\documentclass[11pt]{cernyrep}
\usepackage{graphicx,epsfig}
\graphicspath{{./Pics/}}
\usepackage{amsmath}
\usepackage{amssymb}
\usepackage{wasysym}

\usepackage{cite}

\def\Refs#1{Refs.~\cite{#1}}
\def\Ref#1{Ref.~\cite{#1}}
\def\Eq#1{Eq.~(\ref{#1})}

\newcommand{\QTY}[2]{\mbox{\(#1\rm\,#2\)}}

\begin{document}

\title{Atmospheric Neutrinos
\thanks{Prepared for the forthcoming book 
``Particle Physics with Neutrino Telescopes," C. P\'erez de los Heros, editor, (World Scientific)
}}

\author{
  T.K.~Gaisser$^{1}$, 
}

\institute{
{\footnotesize
  $^1$ Bartol Research Institute and Department of Physics and Astronomy, 
  University of Delaware, Newark, DE 19716 USA}
}

\maketitle

\begin{abstract}

Atmospheric neutrinos produced by cosmic-ray interactions around the globe
provide a beam for the study of neutrino properties.  They are also a background
in searches for neutrinos of astrophysical origin.  Both aspects are addressed in
this chapter, which begins with a brief introduction on neutrino oscillations
in relation to the spectrum of atmospheric neutrinos.
Section 2 describes the cascade equation for hadrons in
 the atmosphere and the main features of atmospheric leptons from their decays.
Next, uncertainties in the fluxes
that arise from limited knowledge of the primary spectrum and of
particle production are discussed.  The final section covers aspects
specific to neutrino telescopes.

\end{abstract}
\section{Introduction}\label{sec:introduction}
The discovery of atmospheric neutrino oscillations was announced twenty
years ago by Super-Kamiokande~\cite{Fukuda:1998mi}.  The key observation was the 
energy-dependence of the ratio of electron to muon neutrinos, comparing
fluxes from below the horizon with those from above.
Simple counting of the low-energy decay chains leads to the expectation
of two muon neutrinos for every electron neutrino.  
The observed ratio was closer to one for sub-GeV neutrinos and remained
low into the GeV region for those from below with pathlengths of $\sim 10^4$~km,
while the ratio increased for the neutrinos produced above
the detector at the typical altitude of $15$~km.  The interpretation was that
muon neutrinos and tau neutrinos mix and can oscillate between
each other, and the low-energy $\nu_\tau$ were below detection threshold.

Interpretation of the measurements was based on calculations of the
flux of atmospheric muon- and electron-neutrinos that existed at the 
time~\cite{Barr:1989ru,Agrawal:1995gk} and~\cite{Honda:1990sx,Honda:1995hz} in which
secondary neutrinos were assumed to be in the same direction as the parent
cosmic-rays from which they descended.  The importance of making three-dimensional
calculations for low-energy neutrinos was emphasized by Battistoni~{\it et al.}~\cite{Battistoni:1999at}. 
There followed other three-dimensional calculations that also accounted for
the significant effects of the geomagnetic field at Kamioka, as well as for
solar modulation, for example~\cite{Honda:2004yz,Barr:2004br}.  The flux of atmospheric
neutrinos in the context of oscillations was reviewed at the time in Ref.~\cite{Gaisser:2002jj},
including the important azimuth angular dependence (``East-West effect") that results
from the high geomagnetic cutoffs
in Japan.  Effects of solar modulation were also discussed, and the primary spectrum used 
for the calculations was described.  

Measurements of atmospheric neutrinos at Soudan~\cite{Sanchez:2003rb} 
and MACRO~\cite{Ambrosio:2003yz}
confirmed the Super-K results.  To a good first approximation,
the atmospheric neutrinos are interpreted in a two flavor scenario
characterized by $\theta_{23}\approx\frac{\pi}{4}$ and 
$\delta m^2_{23}\approx 2.5\times 10^{-3}$ eV$^{2}$.  Reference~\cite{Richard:2015aua}
reports the fluxes of atmospheric neutrinos after oscillations and averaged
over directions as observed in the four phases of Super-K operation.  The strong azimuthal
effect is studied in detail and well understood in terms of geomagnetic effects
on the incident cosmic rays.  The effect of solar modulation is more difficult
to see at Super-K because of the relatively high geomagnetic cutoff.
Oscillations of atmospheric neutrinos measured by IceCube~\cite{Aartsen:2017nmd}
in the energy range 6 to 56 GeV show consistency with long-baseline results
at lower $E_\nu$.  

Measurement of both charged current
and neutral current interactions of solar neutrinos by SNO~\cite{Jelley:2009zz} determined that
the solar neutrino problem was the result of oscillations involving electron
neutrinos, characterized by parameters $\theta_{12}\approx 33^\circ$
and $\delta m^2_{12}\approx 7.5\times 10^{-5}$eV$^2$.  
Additional measurements by Daya Bay~\cite{An:2012eh}, RENO~\cite{Ahn:2012nd} and Double Chooz~\cite{Abe:2012tg} 
reactors using electron anti-neutrinos from reactors
determined $\theta_{13}$, the third link in the three-flavor oscillation framework. 
Long baseline measurements, KamLAND~\cite{Ahn:2006zza}, MINOS~\cite{Adamson:2013whj} and T2K~\cite{Abe:2013fuq},
coupled with reactor experiments provide further refinement of the oscillation parameters.
Reference~\cite{Abe:2017aap} is the three-flavor analysis of the Super-K data.
For a full review of neutrino oscillations see the PDG article~\cite{Agashe:2014kda,Tanabashi:2018abc}.

An important implication of atmospheric oscillations is that there must be a substantial flux
of atmospheric $\tau$-neutrinos transformed via oscillations from muon neutrinos.  These are, however, not easy to
measure because of the small cross section in the threshold region to produce the $\tau$-lepton
with mass~$\approx 1.9$ m$_p$.  Measurements at Super-K~\cite{Abe:2012jj} with atmospheric neutrinos and 
at OPERA~\cite{Agafonova:2018auq} at Gran Sasso in the beam from CERN are consistent with expectation.
An effort to measure $\nu_\tau$ appearance with neutrinos in IceCube DeepCore is underway~\cite{DeYoung:2018abc}.
One goal is to check for unitarity of the three-flavor mixing matrix.  A deficit of $\tau$-neutrinos
could be a signal for sterile neutrinos.  Other possible signals of physics beyond the standard model 
discussed in this volume include non-standard neutrino interactions and violation of Lorentz
invariance.  For all such searches, a good knowledge of the atmospheric neutrino spectrum
is essential.

The main implication for neutrino astronomy is that neutrinos from distant sources
should contain $\tau$-neutrinos at comparable levels to the other two flavors~\cite{Learned:1994wg}.


\section{Phenomenology of neutrino spectra}\label{sec:phenomenology}

The atmospheric cascade is initiated by primary cosmic rays interacting with nuclei in
the atmosphere.  Since production of mesons occurs at the level of nucleon-nucleon interactions,
calculation of the  
inclusive\footnote{"Inclusive spectrum" refers to a measurement of
the rate at which particles of a given type pass through an infinitesimal detection area, 
independent of whether other particles of the same type were produced by the same
primary cosmic-ray.}
 spectra of atmospheric leptons starts from the spectrum of nucleons as a function
 of energy per nucleon.  In the superposition approximation, the 
 points of first interaction of the nucleons from a primary cosmic-ray of mass number $A$
 are distributed on average as if each of its nucleons were injected independently and interacted
 with an interaction length in g/cm$^2$ of
 \begin{equation}
 \label{interaction_length}
 \lambda_N = A_{\rm air}\,m_p\,/\,\sigma_{p-air}^{\rm inel}
\end{equation}
 
 The cascade equation for hadrons in the atmosphere is then
\begin{eqnarray}
\label{MasterEqn}
\frac{{\rm d}N_i(E_i,X)}{{\rm d}X} &=&-\frac{N_i(E_i,X)}{\lambda_i}
                                             - \frac{N_i(E_i,X)}{d_i} \\
& & +\sum_{j=i}^J\int_E^\infty\,\frac{F_{ji}(E_i,E_j)}{E_i}\,
\frac{N_j(E_j,X)}{\lambda_j}\,{\rm d}E_j. \nonumber
\end{eqnarray}
The variable $X$ is the slant depth in g/cm$^2$ of atmosphere along the direction of the
primary cosmic-ray, and $N_i(E_i,X)$ is the number of particles of type $i$
at that slant depth.  Interaction lengths $\lambda_i$ for each type of hadron 
are defined in analogy to the interaction length for nucleons in Eq.~\ref{interaction_length}.
The first term on the right-hand side of Eq.~\ref{MasterEqn} is loss due
to interactions in the atmosphere, while the second represents loss by decay.
For consistency, the decay length has to be expressed in g/cm$^2$, so
\begin{equation}
\label{decay-length}
d_i(X) = \rho(X)\,\gamma\,c\,\tau_i, 
\end{equation}
for a particle with rest lifetime $\tau_i$
and Lorentz factor $\gamma$ at a slant-depth $X$ where the density of the
atmosphere is $\rho(X)$.

\begin{figure}[t]
\centerline{\includegraphics[width=9.cm]{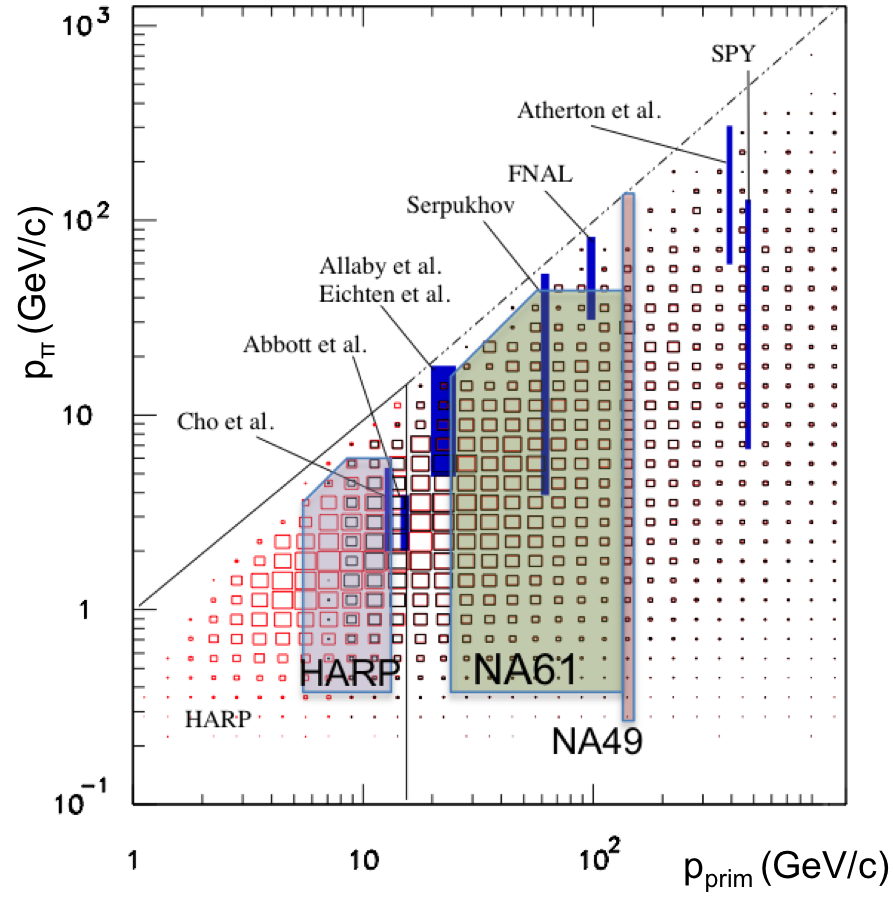}}
\caption{Phase space diagram for the calculation~\cite{Barr:2004br} of contained neutrinos at Super-K.
(Figure updated from Ref.~\cite{Barr:2006it}, see text for explanation.)}
\label{fig:measured}
\end{figure}

The last term in Eq.~\ref{MasterEqn} accounts for production of particles of
type $i$ by particles of type $j$ and energy $E_j> E_i$.
Its definition is
\begin{equation}
F_{ji}(E_i,E_j)\equiv
E_i \frac{1}{\sigma^{\rm air}_j} \frac{{\rm d}\sigma_{j\,{\rm air} \rightarrow i}}{{\rm d}E_i} =
E_i\frac{{\rm d}n_{ji}(E_i,E_j)}{{\rm d}E_i},
\label{inclusive}
\end{equation}
where d$n_i$ is the number of particles of type $i$ produced on average
in the energy bin d$E_i$ around $E_i$ per collision of an incident particle
of type $j$.  Defined in this way, $F_{ji}$ is a dimensionless quantity.
This definition is motivated by the fact that cross sections vary slowly
with energy and the inclusive cross sections at high energy exhibit
an approximate scaling behavior such that $F(E_i,E_j)\approx F(x_L)$,
where $x_L = E_i/E_j$ is the ratio of lab energies.
Spectrum-weighted moments of $F(x_L)$ defined as
\begin{equation}
\label{Z-factors}
Z_{ji}\equiv\int_0^1\,(x_L)^{\gamma-1}\,F_{ji}(x_L)\;{\rm d}x_L\,
=\,\int_0^1 (x_L)^\gamma\frac{{\rm d}n_{ji}}{{\rm d}x_L}{\rm d}x_L
\end{equation}  
appear in the approximate solutions to Eq.~\ref{MasterEqn} to be discussed below,
in which the spectrum of nucleons is approximated as a power law with
an integral index $\gamma$. 

The blue bars in Figure~\ref{fig:measured} from Ref.~\cite{Barr:2006it} show the phase space
covered by experiments used for the calculation of Ref.~\cite{Barr:2004br}.
More recent data sets from HARP~\cite{Catanesi:2008zzc}, NA49~\cite{Alt:2006fr,Anticic:2009wd} 
and NA61/SHINE~\cite{Abgrall:2013qoa,Aduszkiewicz:2017sei} are also indicated.  The underlying 2-D histogram shows the distribution of the
primary cosmic-ray nucleon and parent pion energies for neutrinos in the energy region of contained event sample
of Super-K from the simulation of Ref.~\cite{Barr:2004br}.  The red area
below $8$ GeV is excluded by the geomagnetic cutoff at Super-K.  Each row in the histogram is
the integrand of $Z_{p\pi}$ from Eq.~\ref{Z-factors} weighted by the efficiency for ``contained" events.
The integrand of the Z-factor is proportional to the product of the spectrum of nucleons and the yield of charged pions.
The distribution for a higher-energy event sample would be shifted to higher energy, but would
display the same qualitative features of a rise of intensity at threshold and decrease at high energy
reflecting the steepness of the primary spectrum.  These factors are illustrated below in 
Fig.~\ref{fig:response}.  Estimates of uncertainties in the meson
yields in Ref.\cite{Barr:2006it} need to be updated to include the more recent data sets and to extend the uncertainty
estimates to higher energy.

For unstable mesons the competition between decay and interaction is determined 
by Eq.~\ref{decay-length} compared to the interaction length.  The pressure
at vertical depth $X_v$ is determined by the weight of the column of air above; 
$P=g X_v\approx gX\cos\theta$, where the flat Earth approximation for
slant depth is good for zenith angle $\theta< 70^\circ$.  Using the
ideal gas law to relate density, pressure and temperature leads to the relation
\begin{equation}
\label{critical-energy}
\frac{1}{d_i} = \frac{\epsilon_i}{EX\cos\theta}\,\,{\rm with}\, 
\epsilon_i=\frac{m_ic^2}{c\tau_i}\frac{RT}{gM},
\end{equation}
with $M=0.02896$~kg/mol for air.  The quantity $RT\,/\,gM$ has
the dimension of length.  It is the local scale height at $X_v$
of an exponential approximation to the atmospheric density
as a function of altitude:
$X_v \approx X_0\exp\{-h/h_0\}$ with $h_0\approx 6.5$~km for a
temperature of $220^\circ$K and $X_0\approx 1030$	g/cm$^2$ at sea level.  
For example, when $E_\pi\approx \epsilon_\pi$
the decay and interaction terms in Eq.~\ref{MasterEqn} are comparable
when $X_v = \lambda_N\approx 85$~g/cm$^2$, which corresponds to an
altitude of about $15$~km, the typical production height of most muons.

For $E_i << \epsilon_i$ the parent mesons decay before interacting
and the lepton spectrum directly reflects the primary spectrum
of nucleons.  As energy increases ($E_i> \epsilon_i$), decay
becomes increasingly unlikely relative to re-interaction of the
mesons.  In the high-energy limit, the meson fluxes reflect
the primary spectrum, and their probability of decay is proportional
to the decay probability.  For example, the spectrum of decaying
pions is
\begin{equation}
{\cal D}_\pi(E,X)\,=\,\frac{1}{d_\pi}\Pi(E,X)\,=\,\frac{\epsilon_\pi\sec\theta}{EX}\Pi (E,X),
\label{Dpi}
\end{equation}
where $\Pi(E,X)$ is the solution of Eq.~\ref{MasterEqn} for pions in the high-energy
limit.  Thus the spectrum of atmospheric leptons at high energy is asymptotically
one power steeper than the parent hadron spectra.  The $\sec\theta$ factor in Eq.~\ref{Dpi}
implies that the transition to the steeper energy spectrum occurs at higher
energy for large zenith angles.  As a consequence, for a given $E > \epsilon$,
the intensity of leptons increases with increasing zenith angle.

It is worth noting that the basic features of the spectrum of atmospheric neutrinos
were explained by Zatsepin and Kuz'min~\cite{Zatsepin:1961abc} in 1961,
including the important role of kaons relative to pions
as parents of neutrinos.  Their papers include the production of
neutrinos from decay of muons, and they describe the strong angular dependence
at high energy, which is a consequence of the $\sec\theta$ dependence
of the ratio of decay to interaction above the critical energies of
the mesons.  They also explain that charged kaons are more efficient
producers of neutrinos than charged pions because of the higher mass of
the kaon relative to the muon and the shorter lifetime of the kaon.  (See 
the discussion of Eq.~\ref{decay-factors} below.)
In the 60's and 70's Volkova and Zatsepin published a series of papers
refining the early calculations, as described in the paper of Volkova~\cite{Volkova:1980sw},
which remains a standard reference for fluxes of atmospheric neutrinos.

\subsection{Solving the cascade equations}
\label{subsec:solving}

The most general approach to solving the cascade equations is a full Monte Carlo.
This method is necessary for atmospheric neutrinos at low energy where three-dimensional
effects must be accounted for.  It is also the standard approach for air showers.
Each event generated is an instance of a solution of the cascades equations subject
to a $\delta$-function boundary condition, namely $N_A(E,X=0)=\delta(E-E_A)$ for
the primary nucleus of mass $A$ and zero for all other species.  Simulated events
need to be re-weighted to match a specific primary spectrum and composition.  However,
a full Monte Carlo becomes inefficient for atmospheric leptons 
of high energy because the decay probability of the parent mesons decreases with energy according to Eq.~\ref{critical-energy}.
Biased Monte Carlos may be designed to increase the efficiency for special purposes
by rejecting events as soon as it can be ascertained that they will not
be included in the desired class of events.  A recent example is the 
generalized atmospheric-neutrino self-veto~\cite{Arguelles:2018awr} where 
specialized simulations~\cite{Jero:2016abf}
are used to verify the calculation.

For the calculation of atmospheric muons and neutrinos above a few GeV, 
linear solutions of the cascade equations are sufficient. 
In this case, a classic approach is to integrate
the equations numerically step by step through the atmosphere.  The most advanced
and modern version of this approach is the Matrix Cascade Equation (MCEq) 
solver~\cite{Fedynitch:2016nup,Fedynitch:2017abc}.  
MCEq has been used recently~\cite{Fedynitch:2018cbl} to obtain fluxes of atmospheric
muons and neutrinos from Sibyll~2.3c~\cite{Riehn:2017mfm}.  An advantage of MCEq is that all
relevant hadronic species are followed so that relatively small contributions
to the fluxes can be assessed, including, for example, the contribution of the
decay of unflavored mesons to prompt muons.  In addition, the solver handles
the integration along trajectories near the horizon correctly.

To demonstrate the basic features of the energy- and angular-dependence of
atmospheric neutrinos, approximate analytic solutions of Eq.~\ref{MasterEqn}
in which only the main contributing mesons are included are instructive.
Three simplifying assumptions lead to the solutions:
\begin{enumerate}
\item a power-law in energy for
the differential spectrum
of primary nucleons\\ $\phi_N(E_N)\propto E_N^{-(\gamma + 1)}$,
\item scaling for the inclusive production cross sections\\ 
$F_{ij}(E_i,E_j)=F_{ij}(E_i/E_j)$, and
\item constant interaction and production cross sections.
\end{enumerate}
The boundary condition for inclusive fluxes is 
$N_i(E,X=0) = \phi_N(E_N)$, where $\phi_N$ is the 
differential flux of nucleons, and zero for all other hadrons.

Analytic solutions are obtained for the production
spectrum of leptons as a function of slant depth $X$ in
two energy regions, $E\cos\theta$ much less
than and much greater than the relevant critical energies.
For example, for $E_\pi\cos\theta >> \epsilon_\pi$ the
production spectra of $\nu_\mu$ and $\mu$ are obtained from Eq.~\ref{Dpi}
by multiplying by the decay distribution for 
$\pi^\pm\rightarrow \mu +\nu_\mu$ and integrating over the
parent pion energies allowed for a given lepton energy.
In the center of mass system, the momenta of the neutrino
and the muon are equal and opposite, but because of the large
mass of the muon compared to that of the neutrino, the two distributions
are significantly different when transformed to the lab system.
The multiplicative decay factors at high energy are
\begin{equation}
\label{decay-factors}
Z_{\pi\mu}=\frac{(1-r_\pi^{\gamma+2})}{(\gamma+2)(1-r_\pi)}\,\,\,\,{\rm and}\,\,\,
Z_{\pi\nu_\mu}=\frac{(1-r_\pi)^{\gamma+2}}{(\gamma+2)(1-r_\pi)},
\end{equation}
where $r_\pi = \frac{m_\mu^2}{m_\pi^2}=0.573$ and $\gamma \approx 1.7$ is
the integral spectral index.  Numerically, these factors
are quite different for $\nu_\mu$ and $\mu$.  The differences are much
less for the channel $K^\pm\rightarrow \mu +\nu_\mu$ because $r_K\approx0.046$
is small.

Full details of the derivation of the analytic approximations are given in
Chapter~6 of Ref.~\cite{Gaisser:2016uoy}.  Integrating over the production spectra and 
combining the high- and low-energy expressions with an interpolation formula
leads to the following expression for the muon neutrino spectrum ($\nu+\overline{\nu}$):
\begin{eqnarray}
\label{numuflux}
\frac{{\rm d}N_\nu}{{\rm d}E_\nu} & \simeq & \frac{N_0(E_\nu)}{1-Z_{NN}}\left\{
\frac{{\cal A}_{\pi\nu}}{1+{\cal B}_{\pi\nu}\cos{\theta}\,E_\nu/\epsilon_\pi}
\right.\\ \nonumber
& &\left.+\;
0.635\frac{\,{\cal A}_{K\nu}}{1+{\cal B}_{K\nu}\cos{\theta}\,E_\nu/\epsilon_K}\,+\,
\sum_i B_{D_i}\frac{{\cal A}_{D\nu}}{1 + {\cal B}_{D\nu}\cos{\theta}\,E_\nu/\epsilon_D}\right\}.
\end{eqnarray}
The scaling assumption, in combination with the power-law form for the primary spectrum,
allows the neutrino spectrum to be expressed as a product of the spectrum
of primary nucleons evaluated at the energy of the lepton and a sum of contributions 
from pion decay, from kaon decay and
from decay of charmed hadrons.  The factor $1/(1-Z_{NN})$ accounts for the
regeneration of nucleons in the cascade. 
The contribution from muon decay,
which becomes important at low energy, is not included here.  

The equation 
for $\mu^+ +\mu^-$ has the same form but with different decay factors as
in Eq.~\ref{decay-factors}.  In addition, there is a multiplicative factor
to account for survival probability against energy loss and decay.
  The losses are negligible in the TeV
range and above, but become increasingly significant at lower energies and
for large zenith angles.

\begin{table}[ht]
\caption{Critical energies, $\epsilon_i$ (GeV)}\label{tab:crit-E}
\centering
\begin{tabular}{ccccc} \hline
$\mu$ & $\pi^{\pm}$ & K$^{\pm}$ &  D$^{\pm}$ & D$^0$ \\ \hline
1. & 115. & 850. & 3.9$\times 10^7$ & 9.9$\times 10^7$ \\ \hline
\end{tabular}
\end{table}

\begin{figure}[t]
\centerline{\includegraphics[width=8.cm]{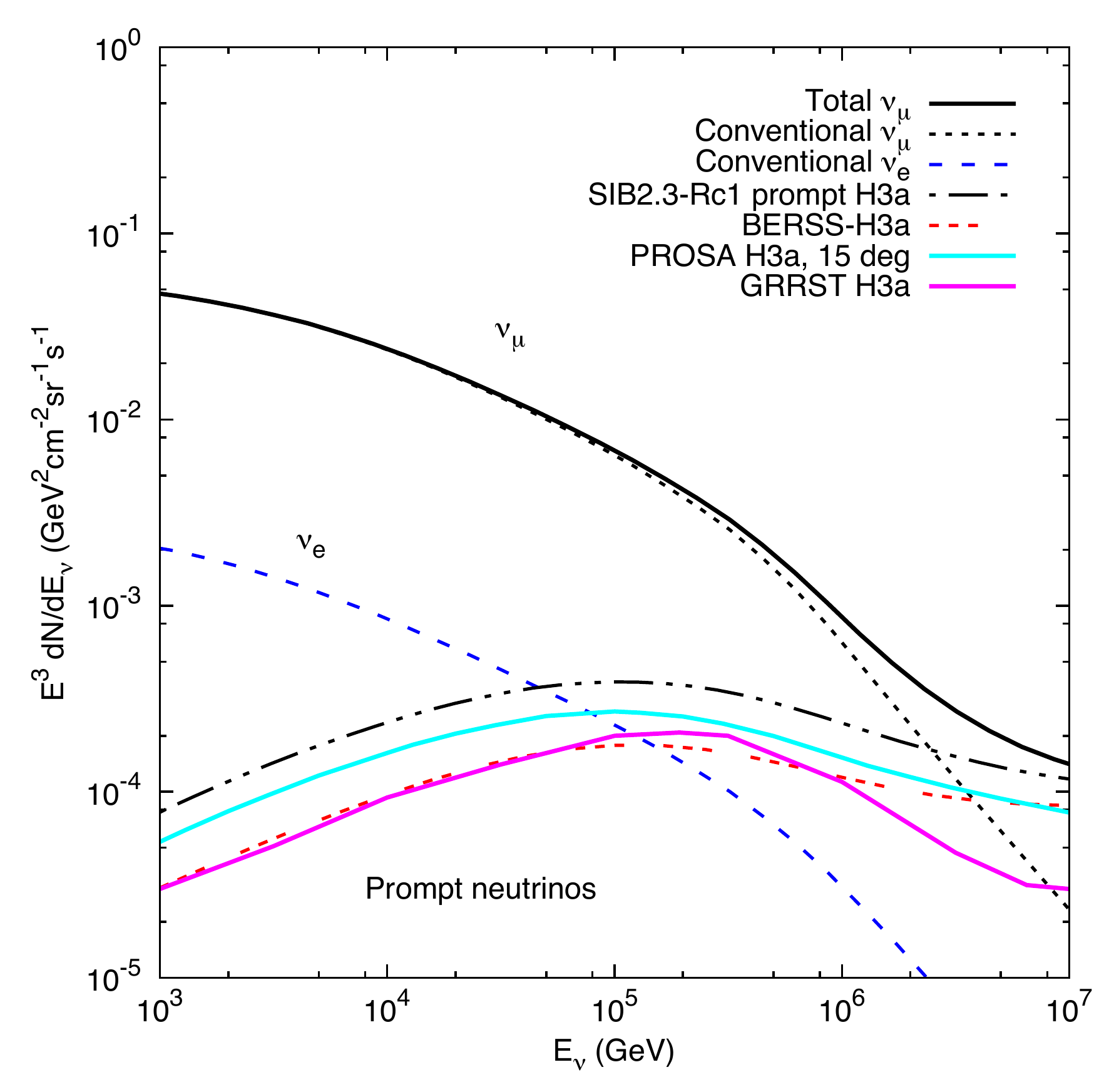}}
\caption{Fluxes of neutrinos ($\nu_\mu+\overline{\nu}_\mu$) with the contribution from charm shown
separately for four models~\cite{Riehn:2015oba,Bhattacharya:2015jpa,Garzelli:2016xmx,Gauld:2015kvh}.  
The flux of electron neutrinos is calculated in Ref.~\cite{Gaisser:2014pda}}
\label{fig:nufluxes}
\vspace{.5cm}
\centerline{\includegraphics[width=9.cm]{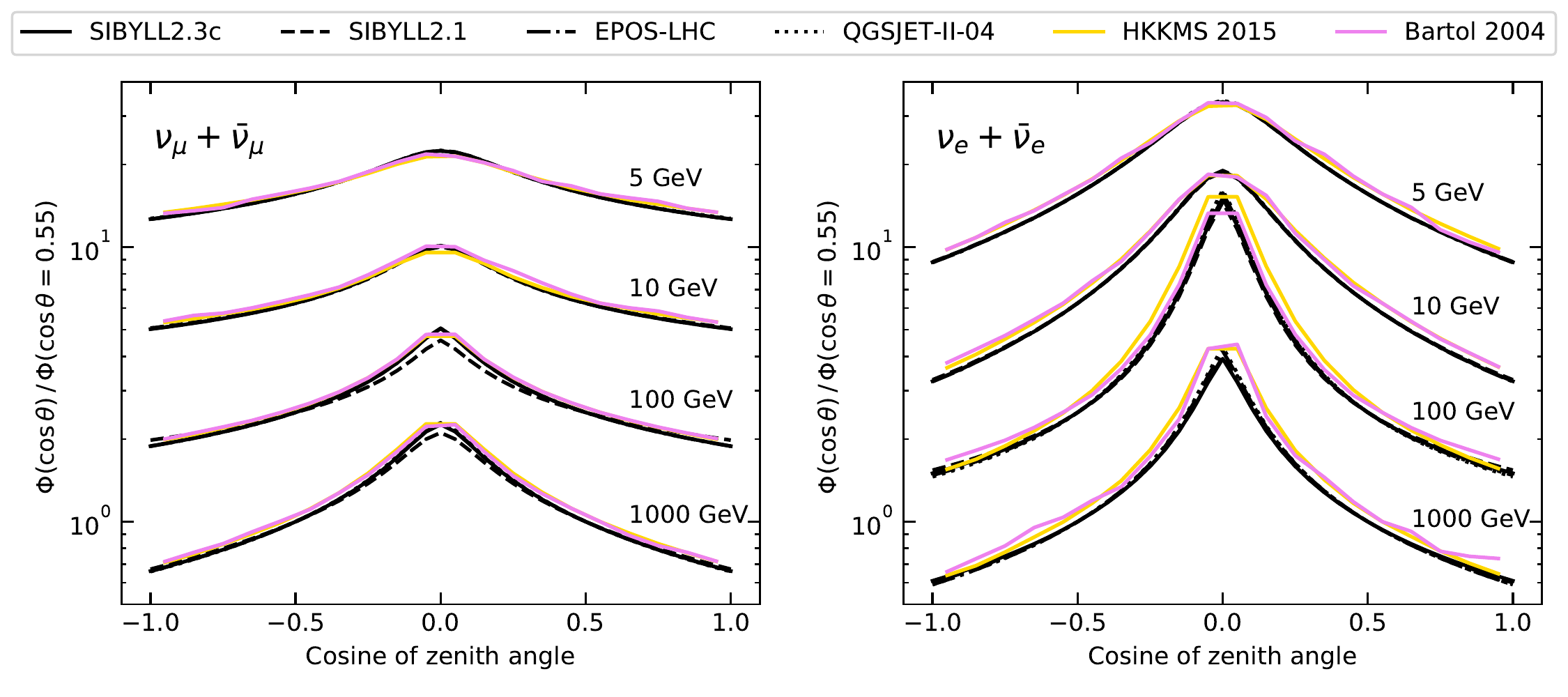}}
\caption{Zenith angle distributions from Ref.\cite{Fedynitch:2018cbl} 
for $\nu_\mu+\overline{\nu}_\mu$ (left) and
$\nu_e+\overline{\nu}_e$ (right) normalized to 1 at $\cos\theta=0.55$.  The lower
energy curves are raised by successive factors of five.}
\label{fig:zendist}
\end{figure}

The basic structure is the same for each term in Eq.~\ref{numuflux}, 
but they contribute differently in different regions of energy and angle
because of their different critical energies (see Table~\ref{tab:crit-E}).
For $E_\nu\cos\theta << \epsilon_i$ the contribution to the neutrino
flux is isotropic and
has the same spectral index as the primary cosmic rays (a differential
spectral index $\approx 2.7$ below the knee).  For $E_\nu\cos\theta >> \epsilon_i$
each contribution to the neutrino flux is proportional to $\sec\theta$
and has a spectral index one power steeper than the primary spectrum.
At each zenith angle there is a gradual steepening of the energy spectrum, which 
occurs at significantly higher energy near the horizontal.  Figure~\ref{fig:nufluxes}
shows the conventional neutrino spectra averaged over zenith angle and
  Fig.~\ref{fig:zendist} from Ref.\cite{Fedynitch:2018cbl} shows
the shapes of the distributions as a function of zenith angle.  The composite energy spectrum
steepens gradually, at first because of the evolution of the angular distribution and then
as a result of the knee in the primary spectrum.  

  The peak of the angular distribution 
near the horizontal for $\nu_\mu$ becomes increasingly prominent as the
vertical fluxes begin to be suppressed by the increasing interaction probabilities
of the parent mesons.  This progression continues as shown
in the left panel of Fig.~\ref{fig:fractional} below.  For electron neutrinos the progression
from $5$ to $100$~GeV reflects mainly the suppression of muon decay near the 
vertical ($\epsilon_\mu\approx 1$~GeV).  
The calculations in Fig.~\ref{fig:zendist} are done with MCeQ~\cite{Fedynitch:2017abc}.
for five event generators and compared to the Bartol 2004 calculation of Ref.~\cite{Barr:2004br}.

\begin{figure}[th]
\centerline{\includegraphics[width=9.cm]{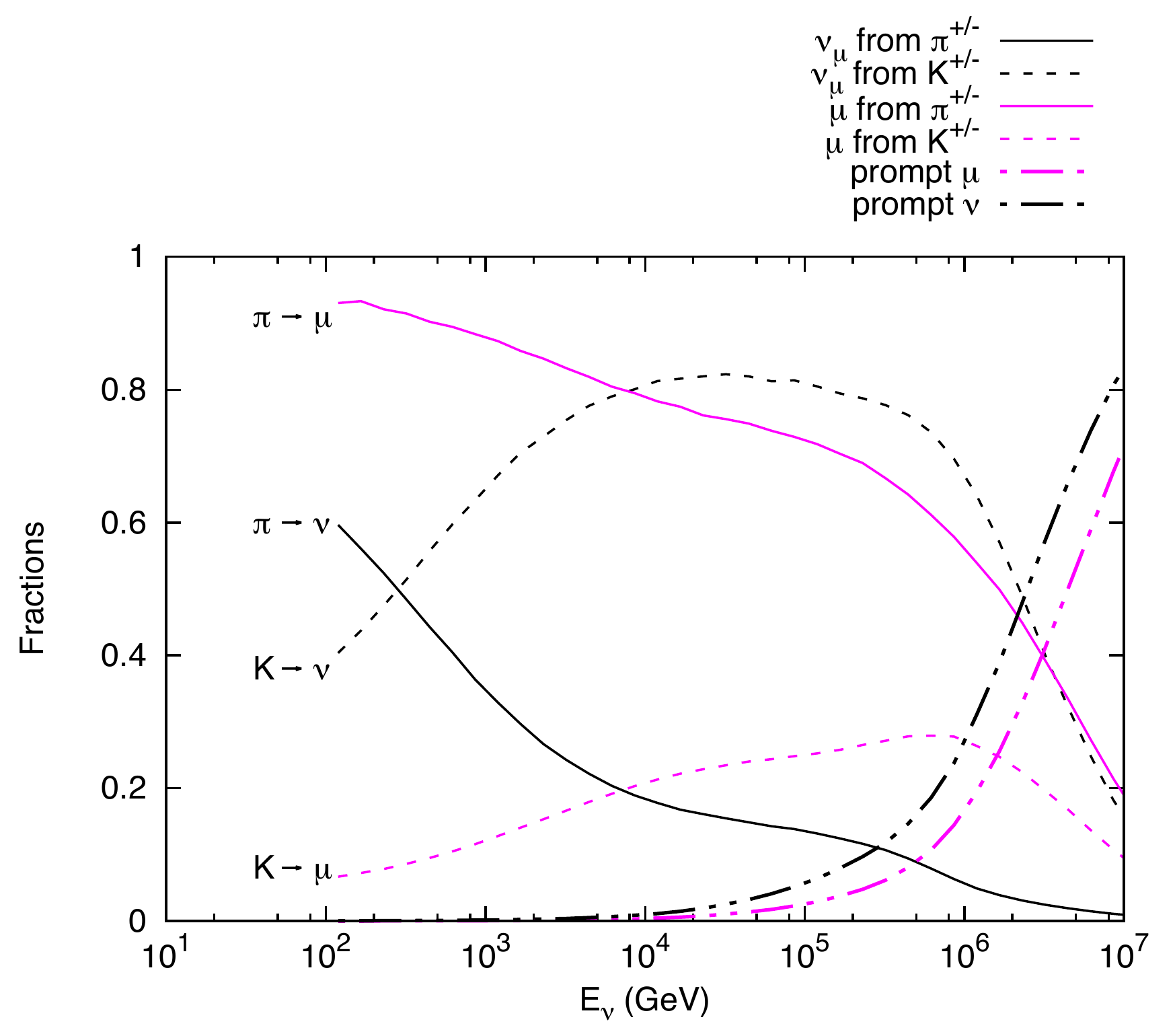}}
\caption{Relative contributions of pions, kaons and charmed hadrons to the fluxes
of muon neutrinos compared to those for muons.  See text for discussion.}
\label{fig:ratios}
\end{figure}

Each contribution in Eq.~\ref{numuflux} depends on the branching ratio for meson decay to $\nu_\mu$
and the spectrum weighted moments that characterize the meson production.
For the pion contribution, for example,
\begin{equation}
\label{factors}
{\cal A}_{\pi\nu}= Z_{N\pi}\,\frac{( 1-r_\pi)^{\gamma+1}}{(1-r_\pi)(\gamma +1)}
\,\,\,{\rm and}\,\,
{\cal B}_{\pi\nu}\equiv{\gamma+2\over\gamma+1}\,{1\over 1-r_\pi}\,
{\Lambda_\pi -\Lambda_N\over \Lambda_\pi\,\ln{(\Lambda_\pi/\Lambda_N)}}.
\end{equation}
The attenuation lengths, $\Lambda_i$, which appear at high energy account for
regeneration of interacting particles of type $i$.  For example, 
$\Lambda_N=\lambda_N/(1-Z_{NN})$.
The decay factor $(1-r_\pi)^{\gamma+1}$ suppresses the pion contribution to $\nu_\mu$
relative to muons, for which the corresponding factor is $(1-(r_\pi)^{\gamma+1})$,
and the suppression is greater at high energy where the exponent becomes $\gamma + 2$.
The effect is much smaller for the kaon channel since $r_K<< r_\pi$ and the neutrino
carries a larger fraction of the parent meson energy than for the pion channel.
In addition, the fact that $\epsilon_\pi^\pm <\epsilon_K^\pm$ means that the pion channel
steepens before the kaon channel.  As a consequence, in the TeV range and above, most $\nu_\mu$ come from
decay of $K^{\pm}$, as illustrated in Fig.~\ref{fig:ratios}.  

Charmed hadrons have much shorter lifetimes, so their critical energies are
more than $10$~PeV and their spectrum at high energy is one power
less steep than the ``conventional" neutrinos from decay of pions and
kaons.  Therefore, despite their small production cross sections,
they eventually become the dominant source of atmospheric neutrinos, as shown
by the lines for prompt neutrinos ($\nu_\mu+\overline{\nu}_\mu$) in 
Fig.~\ref{fig:nufluxes}.\footnote{Charm fluxes in the figure are
calculated for near vertical directions, but they are nearly isotropic for $E_\nu< \epsilon_D$. 
The plotted lines show only the central values of the calculations, which typically~\cite{Garzelli:2016xmx,Gauld:2015kvh}
have theoretical uncertainties comparable to their differences.} 
The yields of $nu_e$, $\nu_\mu$ and $\mu$ are approximately equal in decays of charmed mesons.
However, the prompt muon flux 
includes an additional contribution from decay of unflavored 
mesons~\cite{Illana:2010gh,Fedynitch:2018cbl}.  Nevertheless, the prompt \textit{fraction} for muons
is lower than that for $\nu_\mu$ because the conventional muon flux is significantly higher
than the conventional $\nu_\mu$ flux.

Electron neutrinos at high energy come primarily from three-body decays of charged
and neutral kaons.  Their flux at high energy is about 5\% of the flux
of muon neutrinos.  Fluxes of $\nu+\overline{\nu}$ averaged over zenith angle
are shown in Fig.~\ref{fig:nufluxes} separately for conventional and for
prompt neutrinos.  Calculation of the flux of conventional atmospheric $\nu_e$  
is reviewed in Ref.~\cite{Gaisser:2014pda}, which includes the very small
contribution from K$_S$ that is analogous to the prompt contribution
because of its short lifetime.  The flux of atmospheric $\nu_\tau$ is much smaller
than $\nu_\mu$ and $\nu_e$.  They come mainly from decay of $D_s$~\cite{Pasquali:1998xf}.


\subsection{Muon charge ratio and $\nu_\mu\,/\,\overline{\nu}_\mu$}
\label{subsec:charge-ratio}
\begin{figure}[t]
\centerline{\includegraphics[width=9.cm]{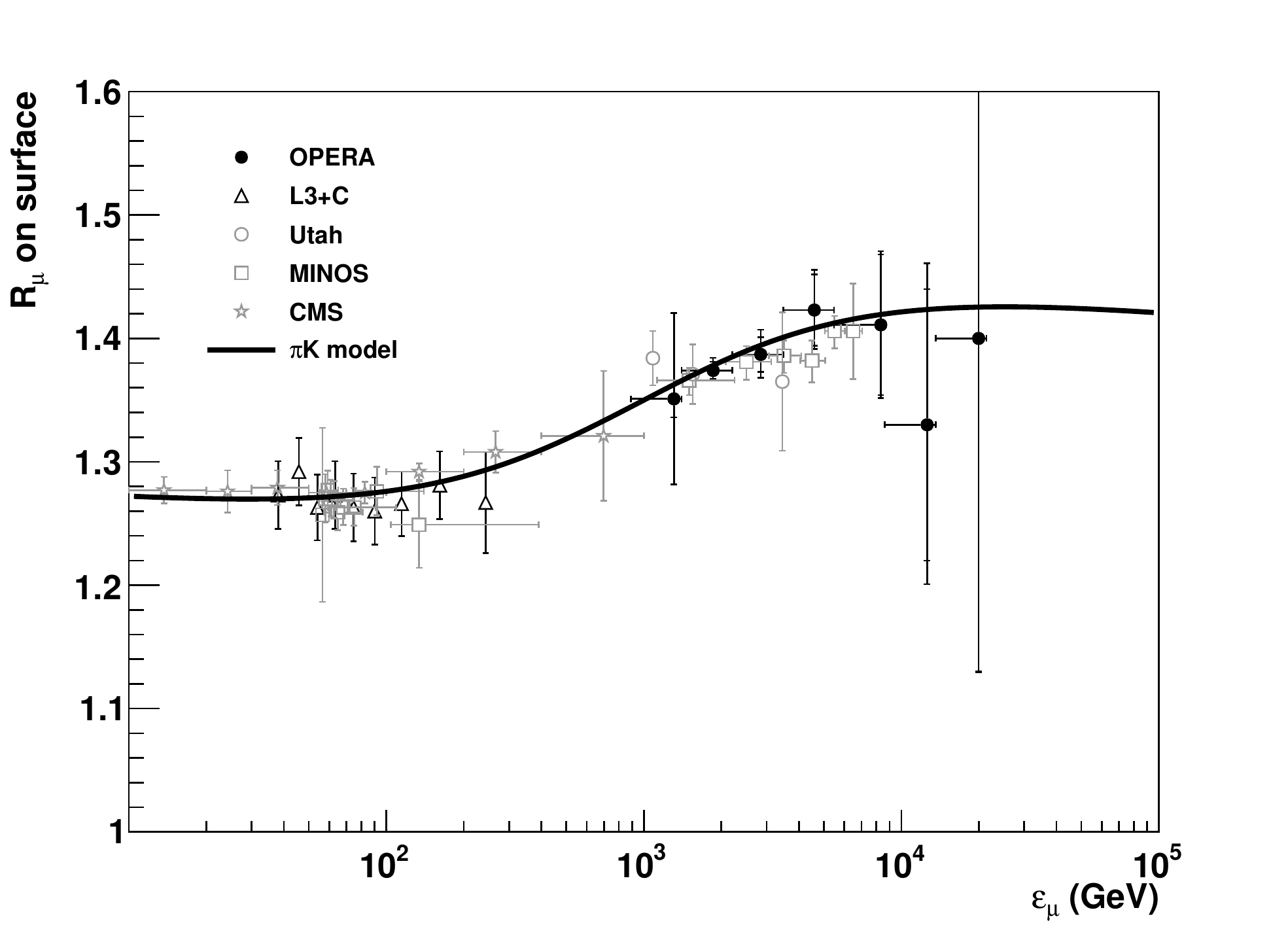}}
\caption{Muon charge ratios as a function of energy.  The figure is 
from OPERA~\cite{Agafonova:2014mzx}.  The line shows their fit to
the data using the parameterization of Ref.~\cite{Gaisser:2011cc}.}
\label{fig:charge-ratio}
\end{figure}

Atmospheric muons are closely related to neutrinos and are much easier to measure.
For this reason, observations of muons can place useful constraints on atmospheric neutrinos.
Indeed, one approach is to calculate the flux of $\nu_\mu$
from measurements of muons at high altitude~\cite{Perkins:1994pm}.  
The Honda group~\cite{Sanuki:2006yd,Honda:2006qj} used measurements of atmospheric muons at various
atmospheric depths to tune
the interaction model of their original three dimensional calculation~\cite{Honda:2004yz}.
A limitation to this approach at TeV energies
is that the neutrinos come much more from the kaon channel, which gives a
relatively small contribution to the muons.  In this connection, it is relevant
to note that there is a feature in
the muon charge ratio that directly reflects the kaon contribution and
can be used to constrain the parameter related to kaon production.  The
$\mu^+/\mu^-$ ratio increases in the energy range approaching a TeV (see Fig.~\ref{fig:charge-ratio}) as the pion contribution steepens 
for $E>\epsilon_\pi\approx 115$~GeV. This increase reflects the associated production channel,
$p\rightarrow \Lambda K^+$, the isospin conjugate of which ($n\rightarrow \Lambda K^0$) does
not produce corresponding $\mu^-$s.

\begin{figure}[ht]
\centerline{\includegraphics[width=7.2cm]{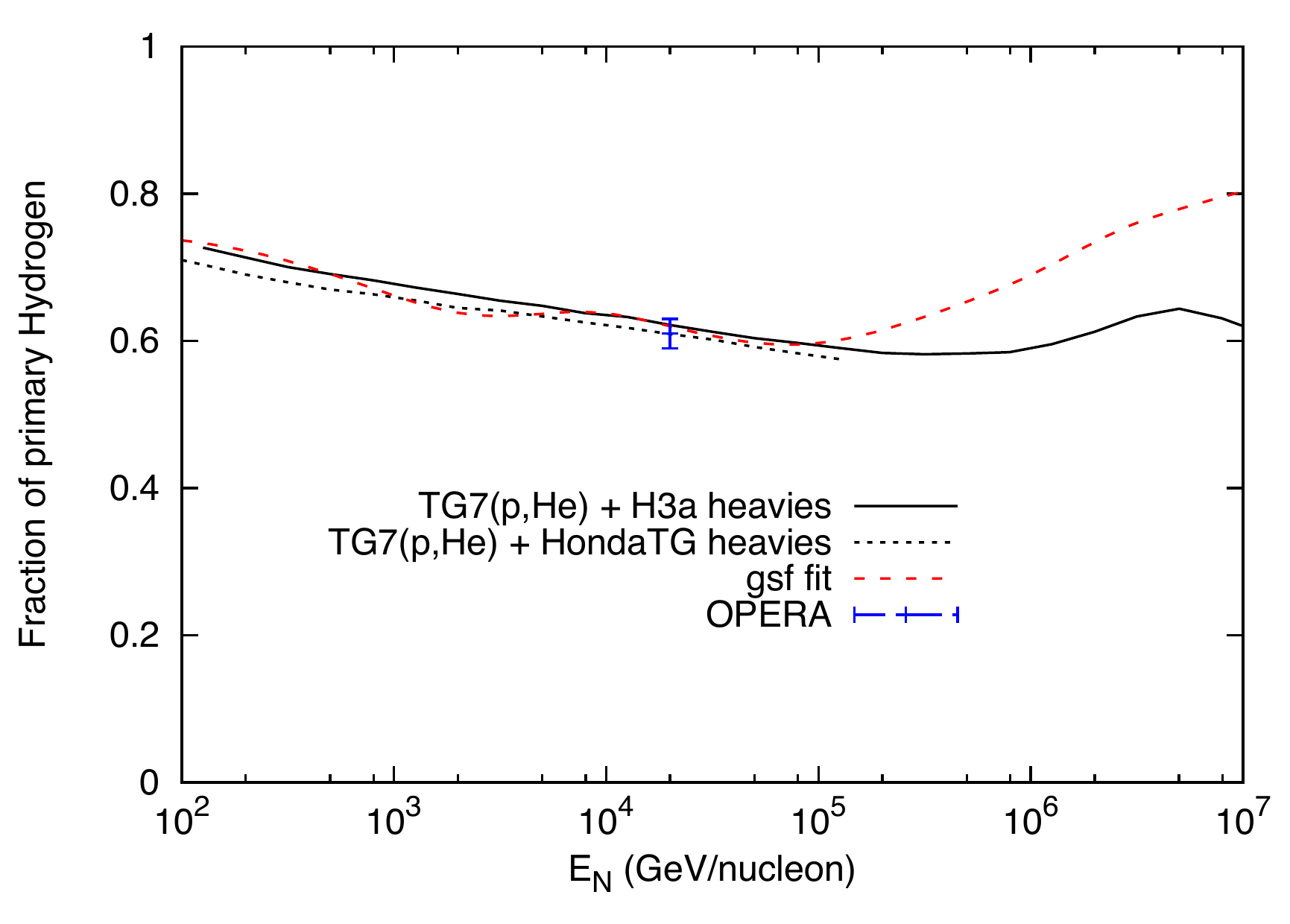}\hspace*{2pt}
\includegraphics[width=7.2cm]{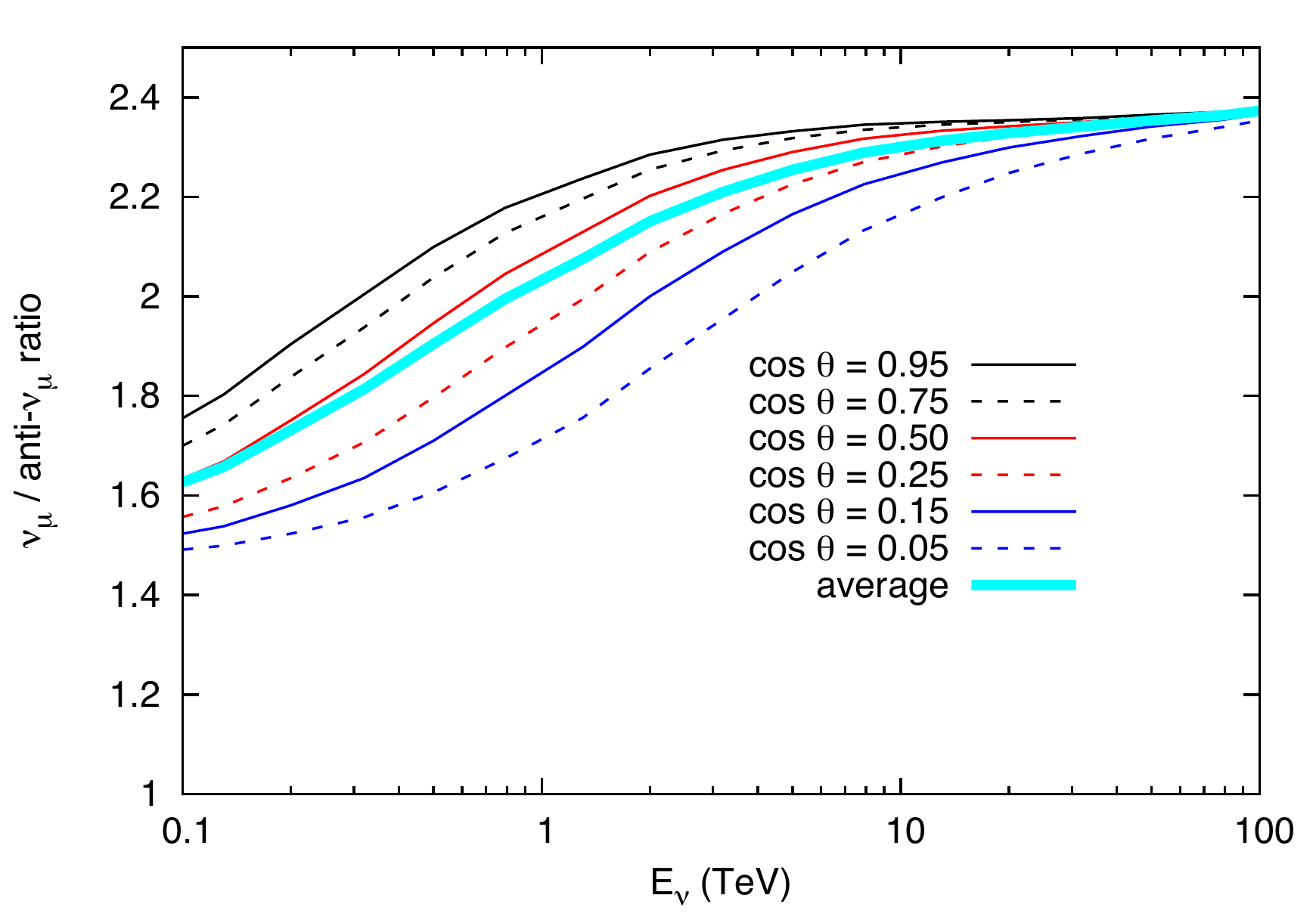}}
\caption{Left: Proton excess in the spectrum of nucleons.  The fitted value of 
  OPERA~\cite{Agafonova:2014mzx} is placed at $20$~TeV primary energy, corresponding
  to $\sim 2$~TeV muon energy.  Spectrum fits are discussed in \S~\ref{sec:primary-spectrum};
  Right: Ratio of $\nu_\mu/\overline{\nu}_\mu$ as a function of energy
and zenith angle calculated with the parameterization of Ref.~\cite{Gaisser:2011cc}
with $\delta_0$ and $Z_{pK^+}$ fit by OPERA~\cite{Agafonova:2014mzx}}
\label{fig:nunu-ratio}
\end{figure}

Calculating the muon charge ratio requires keeping track separately of protons and
neutrons in the primary spectrum and following the charges separately
in solving the cascade Eq.~\ref{MasterEqn}.  The effect of the proton excess in the
primary spectrum of nucleons is contained in the parameter $\delta_0 \equiv (p-n)/(p+n)$,
which decreases with energy up to energies approaching a PeV, as shown in the left panel 
of Fig.~\ref{fig:nunu-ratio}.
For pions, relatively simple solutions for
$\Pi^+(E,X)\pm\Pi^-(E,X)$ can be obtained~\cite{Frazer:1972ep} in terms of 
$\beta\equiv(1-Z_{pp}-Z_{pn})/(1-Z_{pp}+Z_{pn})$ and 
$\alpha_\pi\equiv(Z_{p\pi^+}-Z_{p\pi^-})/(Z_{p\pi^+}+Z_{p\pi^-})\approx 0.17.$
Separate solutions for $K^+$ and $K^-$ are more complicated because the charged kaons
are not isospin conjugates of each other.  In the more general calculation that tracks
 $K^\pm$ separately as well as $\pi^\pm$~\cite{Gaisser:2011cc}, the main parameter for kaons is
$\alpha_K\equiv(Z_{pK^+}-Z_{pK^-})/(Z_{pK^+}+Z_{pK^-})\approx 0.51$.  The larger charge
ratio for kaons shows up in the TeV range as the kaon contribution to the muon flux
increases relative to that of the pions.  The OPERA group fit their data with the
parameterization of Ref.~\cite{Gaisser:2011cc} using 
two free parameters.  They find $\delta_0=0.61\pm 0.02$ and $Z_{pK^+}=0.0086\pm 0.0004$ 
at $E_\mu\approx 2$~TeV, which corresponds to
a primary energy per nucleon higher by roughly a factor of ten.  Their fit to the $\mu^+/\mu^-$
ratio is shown by the line in Fig.~\ref{fig:charge-ratio}.  The OPERA paper also points
out that the prompt contribution to the muon flux has a charge ratio of one
so that the charge ratio will decrease at energies so high that prompt
muons contribute significantly.

Because the kaon channel dominates the production of muon neutrinos, the
$\nu_\mu\,/\,\overline{\nu}_\mu$ ratio is significantly higher than
the charge ratio of muons.  The ratios for muon neutrinos are shown
in Fig.~\ref{fig:nunu-ratio}~(right) with the parameters of\cite{Gaisser:2011cc}
as modified by OPERA.  The evolution to the high-energy plateau occurs faster
for near vertical directions than for the horizontal, reflecting the
$\cos\theta E_\nu\,/\,\epsilon_K$ factor in the denominator of Eq.~\ref{numuflux}.
Different hadronic interaction models predict significantly different
values for the $\nu_\mu\,/\,\overline{\nu}_\mu$ 
ratio~\cite{Fedynitch:2016nup}.  Although all
show the increase with energy associated with the higher charge ratio for kaons,
the magnitudes in the TeV range vary by a factor of two among the models.  
The connection to the measured muon charge ratio is potentially valuable
in this context, but has yet to be fully exploited.

\subsection{Seasonal variations}
\label{subsec:seasonal}
Rates of high-energy muons exhibit a seasonal variation correlated with the
temperature in the upper atmosphere where most of the production occurs.
The effect was first measured and understood in the underground muon
experiment at Cornell~\cite{Barrett:1952abc}.  The variation is a direct
consequence of the fact that the meson decay rate in Eq.~\ref{critical-energy} is proportional
to temperature combined with the fact that the probability for re-interaction before
decay increases with energy.  For example, when $\cos\theta\,E_\mu >>\epsilon_\pi$
in the muon analog of Eq.~\ref{numuflux}, then the d$N_\mu$/d$E_\mu\propto \epsilon_\pi\propto T$.
In this limit, the pion contribution to the muon flux is fully correlated with
the temperature.  For $\cos\theta\,E_\mu<<\epsilon_\pi$, the temperature-dependent term 
is negligible, most mesons decay before they interact and the temperature dependence
disappears for low energy.  There is an intermediate region in the TeV range
where the pion contribution is asymptotic, but the kaon contribution is not, so the
details of the seasonal variation are in principle sensitive to the ratio
of charged pions to kaons in hadronic interactions.  For $E_\mu \le 1$~PeV the
contribution from prompt muons should not depend on temperature because
of the large critical energy of D mesons.

\begin{figure}[th]
\centerline{\includegraphics[width=6.9cm]{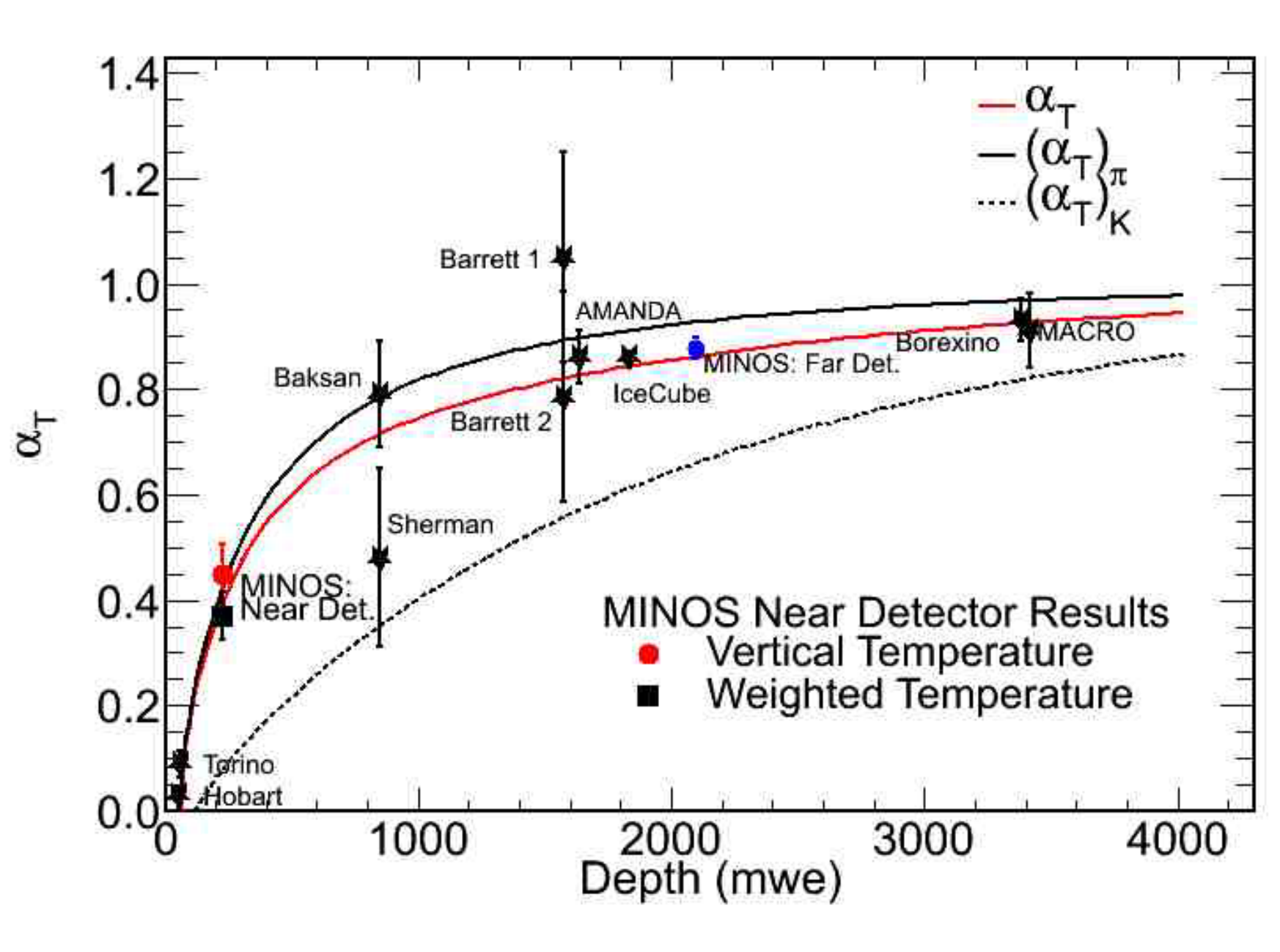}\hspace*{4pt}
\includegraphics[width=6.4cm]{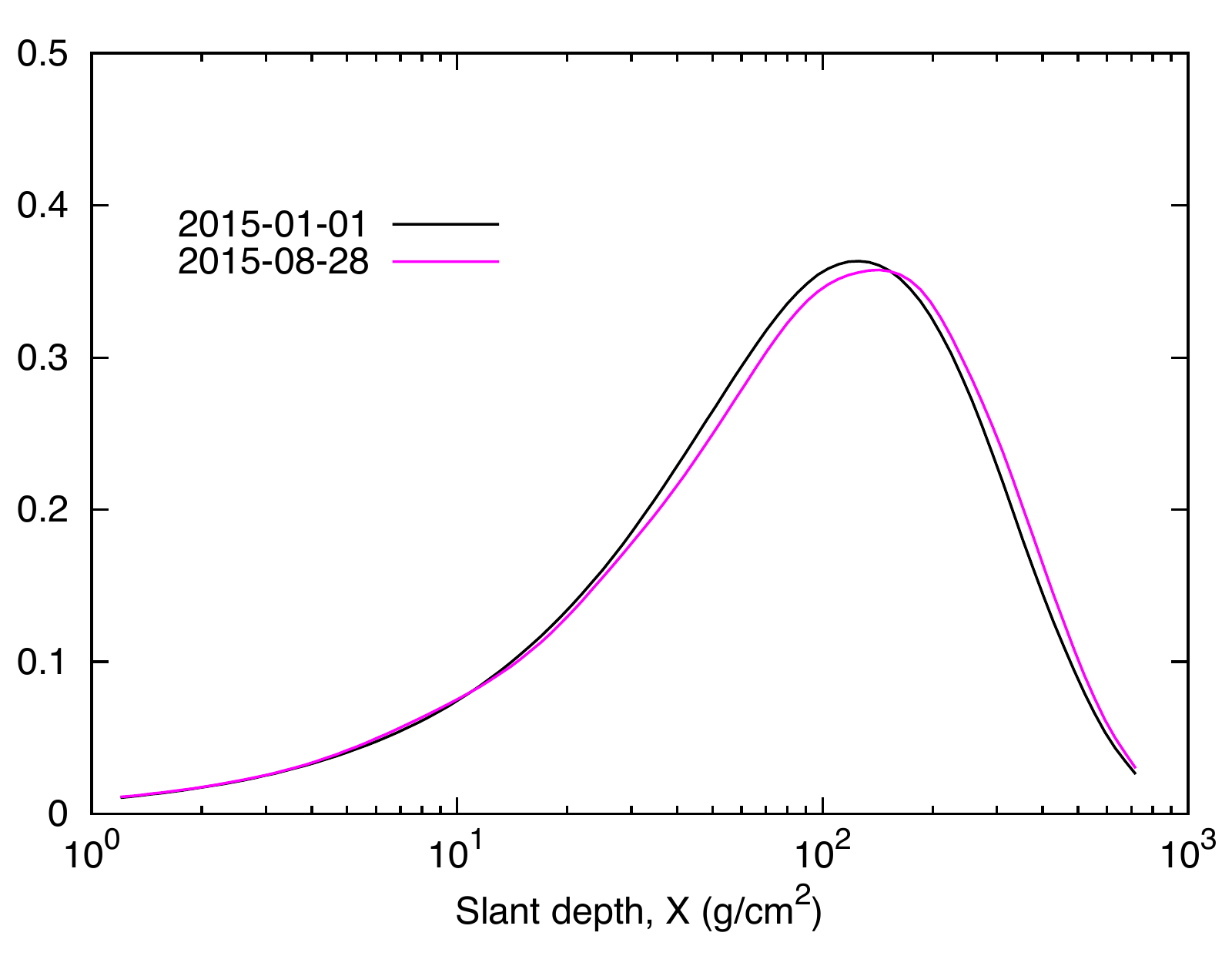}}
  \caption{Left: Summary of measurements of the correlation coefficient between
  muon rate and effective temperature from Ref.~\cite{Adamson:2014xga};
  Right: Normalized production spectrum of muons at a zenith angle of $30^\circ$.}
\label{fig:profile}
\end{figure}

The connection between muon rate and temperature is characterized by a correlation 
coefficient defined in the relation
\begin{equation}\label{alpha_T}
\frac{\delta \phi(E_\mu)}{\phi(E_\mu)}=\alpha_T\frac{\delta T}{T}.
\end{equation}
Following the argument in the previous paragraph, the correlation
coefficient should rise from a small value at low energy and approach
unity at high energy, with a possible decrease in the PeV range
corresponding to the onset of prompt muons~\cite{Desiati:2010wt}.  This expected rise
from shallow detectors ($E_\mu \approx 50$~GeV, at the MINOS near
detector~\cite{Adamson:2014xga}) through IceCube~\cite{Desiati:2011hea} 
and the MINOS far detector at Soudan~\cite{Adamson:2009zf} ($\sim$~TeV) to MACRO~\cite{Ambrosio:1997tc} 
($\sim$~2TeV) is nicely illustrated in the compilation of data in Ref.~\cite{Adamson:2014xga},
reproduced here in the left panel of Fig.~\ref{fig:profile}.

An experimental determination of the correlation coefficient requires
integrating the atmospheric temperature profile weighted by the production
spectrum of muons and the energy and angular dependence of the detector response
to determine an effective temperature,
\begin{equation}
\label{eq:Teff}
T_{\rm eff}(\theta) = \frac{\int{\rm d}E_\mu A_{\rm eff}(E_\mu,\theta)\int{\rm d}X P_\mu(E_\mu,\theta,X) T(X)}
{\int{\rm d}E_\mu A_{\rm eff}(E_\mu,\theta)\int{\rm d}X P_\mu(E_\mu,\theta,X)}.
\end{equation}
Here $A_{\rm eff}$ is the detector acceptance (area x solid angle) for detecting muons of energy $E_\mu$
at production from zenith angle $\theta$, and $P_\mu$ is the production spectrum of muons at slant depth X(g/cm$^2$).
$T_{\rm eff}$ is obtained after averaging over zenith angle.  As an example, the production profile
$P_\mu(E_\mu,\theta,X)$ at 30$^\circ$ is shown in the right panel of Fig.~\ref{fig:profile}.  The muon production
profile itself depends only slightly on the temperature profile (through the temperature dependence
of the critical energy).  The examples in the figure correspond to $T_{\rm eff}=-37^\circ C$
on 2015-Jan-01 and $-81^\circ C$ on 2015-Aug-28 at the South Pole.\footnote{Temperatures 
are from the NASA Atmospheric Infrared Sounding (AIRS) Satellite~\cite{NASA:2018abc}.}

The contributions of pions and kaons are combined in Eq.~\ref{eq:Teff}
and in the red curve in the left panel of Fig.~\ref{fig:profile}.~\footnote{For measurements at low
energy with shallow detectors, it is necessary to account for energy loss and decay of
muons in the atmosphere~\cite{Adamson:2014xga}.}
The contributions for pions only (solid black line) and for kaons only (broken black line)
are also shown separately in the figure.
Because of the relatively higher contribution of kaons to muon neutrinos,
the correlation coefficient for neutrinos is expected to be significantly
lower than that for muons.  A study by IceCube using upward neutrinos from
the partially completed detector\cite{gaisser:2013abc} finds $\alpha_T^\nu\approx 0.5$ with large
uncertainties.  The study uses events with zenith angles $120^\circ \le\theta\le 90^\circ$
corresponding to latitudes between $-30^\circ$ and $-90^\circ$, so the seasonal phase of
the temperature dependence can be compared with that of muons in IceCube produced
in the atmosphere above the South Pole.

\section{Primary spectrum}
\label{sec:primary-spectrum}
Direct measurements of the cosmic-ray spectrum with spectrometers
in space now extend to a rigidity of $\approx 2$~TV, corresponding
to $E\sim$TeV/nucleon.  Measurements
with balloon-borne calorimeters now approach $100$~TeV/nucleon.
Information at higher energies is provided by data from 
extensive air shower (EAS) experiments, which do not directly
detect the primary nuclei.  They generally measure the
all-particle spectrum; that is, the spectrum in terms of energy per particle.
Composition of the primary beam is determined indirectly from
properties of the cascade such as depth of shower maximum and
ratio of electrons/photons to muons in the shower.  The composition
is at best determined to the level of five groups of nuclei:
proton, helium, CNO, Si-Mg and Fe.\footnote{In most cases the
intermediate groups are combined to four groups
with masses that are approximately equally spaced
in $\ln(E)$ from zero to four~\cite{Dembinski:2017zsh}.}  Conversion
to the spectrum of nucleons in terms of energy per nucleon,
which is needed for calculation of fluxes of neutrinos,
therefore becomes more uncertain for $E> 100$~TeV.

For calculations of the flux of atmospheric neutrinos it is useful
to have a parameterization of the primary spectrum anchored to direct
measurements at low energy and guided by the principle of rigidity
dependence to account for the knee of the spectrum.  The Polygonato
Model~\cite{Hoerandel:2002yg} is the classic example, with each and every element
characterized by a normalization and spectral index at low energy
and extrapolated to air shower energies ($\sim PeV$).
A rigidity-dependent change of slope is assumed
to characterize the steepening of the spectrum around $3$~PeV.
The fact that both propagation and acceleration of cosmic rays depend on magnetic
rigidity justifies the assumption that features in the spectrum
should occur at the same rigidity for all nuclei.  Rigidity is $Pc\,/\,Ze$, which is
the total energy per particle divided by its charge.  A consequence
of the rigidity dependence is that the steepening of the different
elements occurs in a sequence known as the Peters Cycle~\cite{Peters:1961abc},
with protons first, followed by He, CNO, etc.  If the knee is characterized
by a rigidity $R_c\approx 4\times 10^6$~GV, then the steepening occurs at
total energy per particle
\begin{equation}
\label{eq:Peters}
E_{tot}^c\,=\,A\times E_{N,c}\,=\,Ze\times R_c,
\end{equation}
where $A$ and $Z$ are the mass and charge of a nucleus with
energy per nucleon $E_N$.

\begin{figure}[t]
\centerline{\includegraphics[width=6cm]{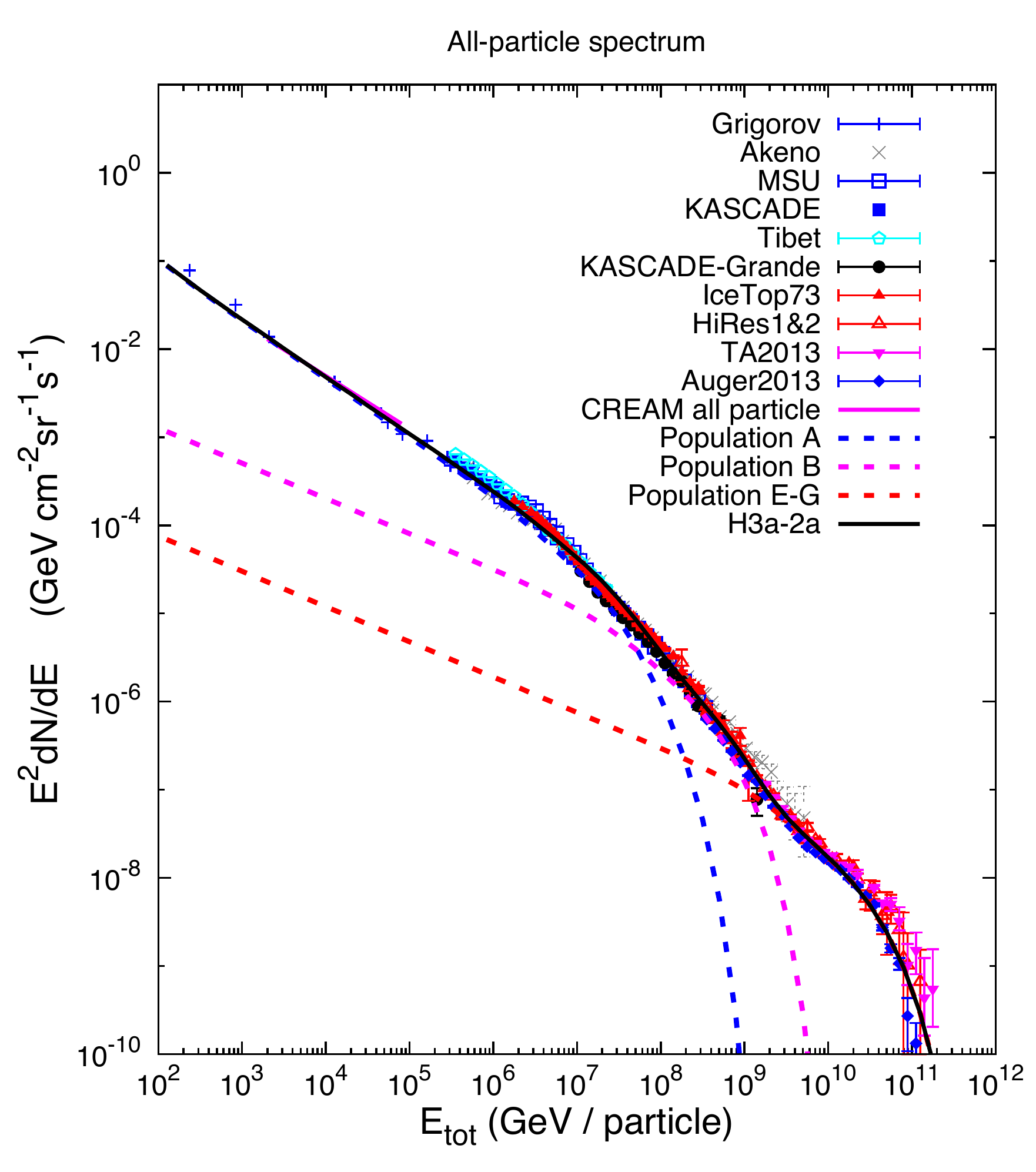}\hspace*{4pt}
\includegraphics[width=6cm]{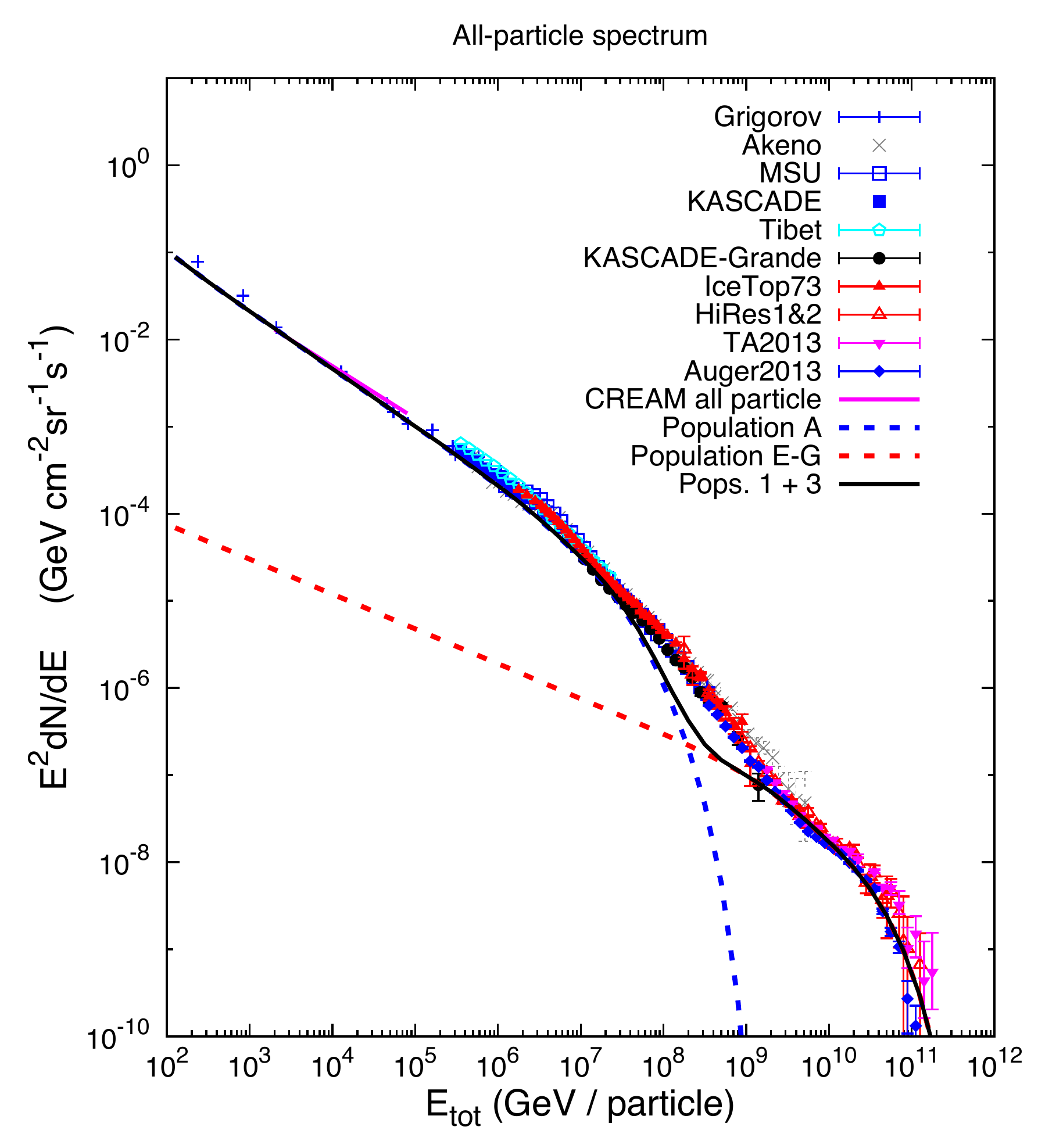}}
  \caption{Left: All-particle spectrum showing separately the contribution of
three populations in a modified H3a model (see text); Right:All-particle 
spectrum with the second Galactic population omitted.}
\label{fig:2pops}
\centerline{\includegraphics[width=6cm]{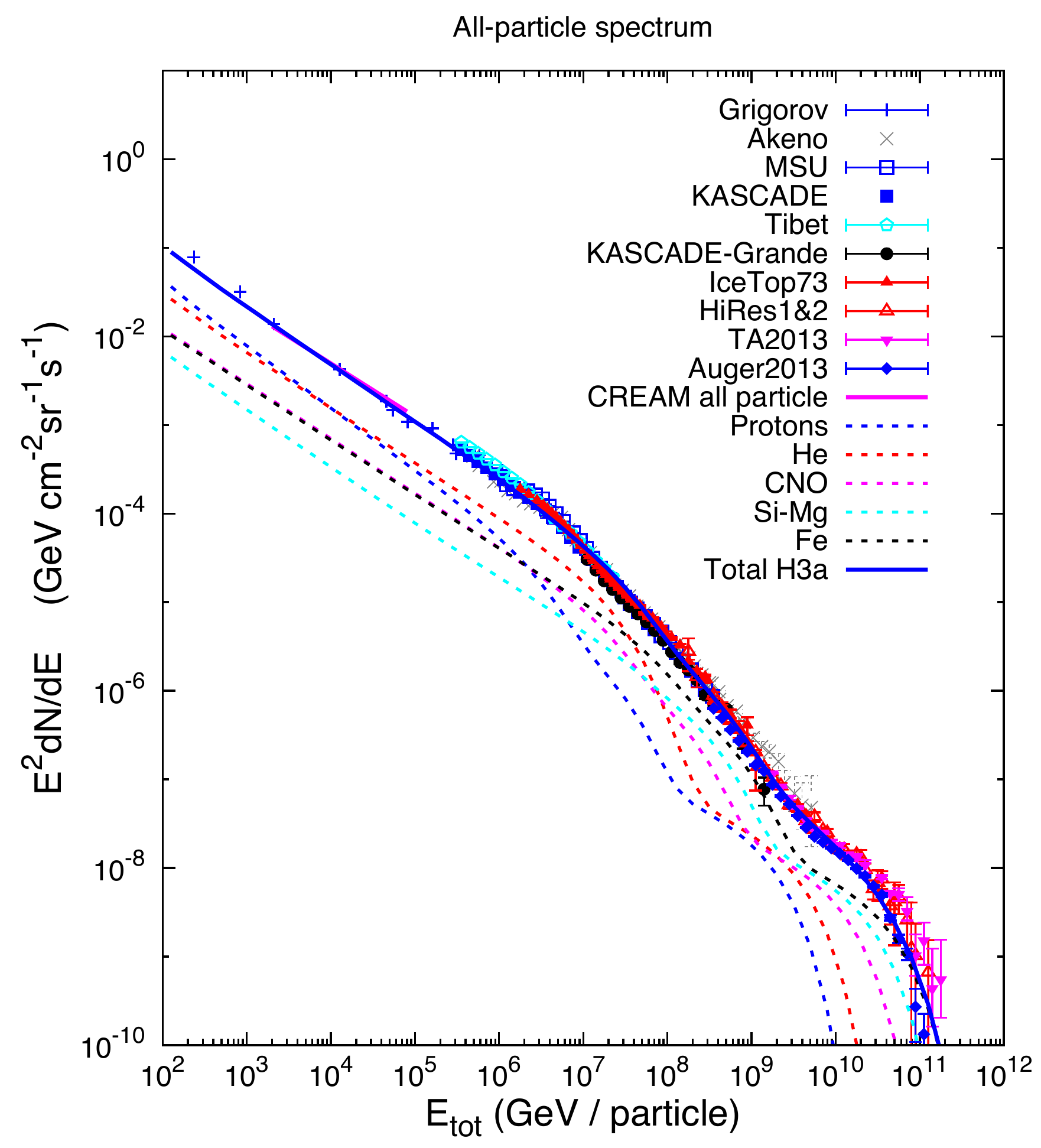}\hspace*{4pt}
\includegraphics[width=6cm]{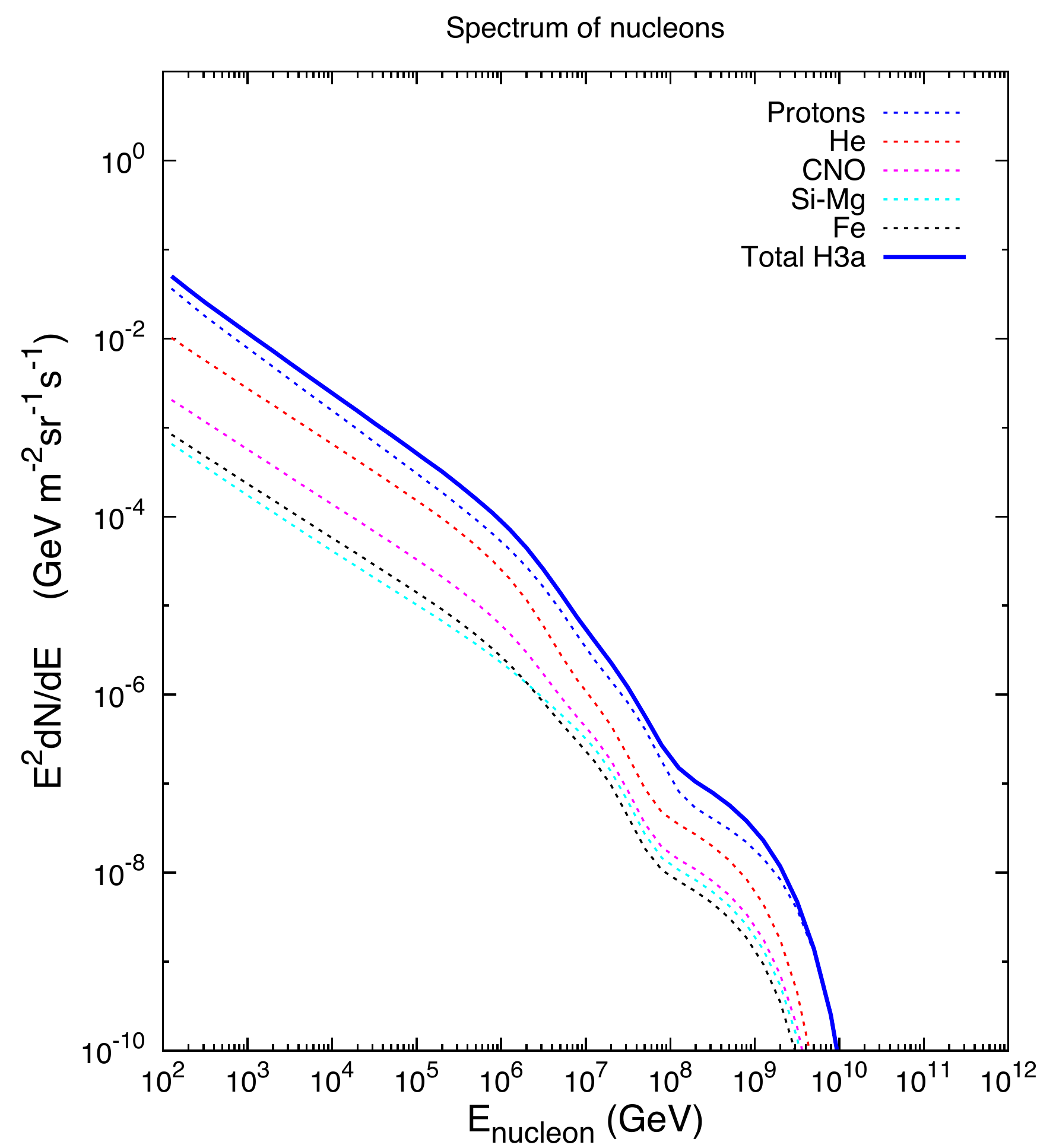}}
  \caption{Left: all-particle spectrum showing separately the contribution of
five mass groups in a modified H3a model (see text); Right: 
spectrum of nucleons corresponding to all-particle spectrum.}
\label{fig:allnucleon}
\end{figure}

A simplified but extended version of the Polygonato Model is 
H3a~\cite{Gaisser:2011cc}.  It is ``simplified" to consist of the standard
five groups of nuclei rather than individual elements, 
and ``extended" to three populations of
particles, a Galactic component that cuts off at the knee, 
a second Galactic component, and an extra-galactic component.
Hillas~\cite{Hillas:2005cs} pointed out that, if the ankle
of the spectrum around $3$~EeV signals the onset of the extra-galactic
population, then a second Galactic component is needed to fill in
the gap after the heavy component of the first population turns
down around $100$~PeV.  Fig.~\ref{fig:2pops}
illustrates the need for a second Galactic component (or a low-energy
extension of the extra-galactic component).

In such a model, the contribution to
the all particle spectrum of nuclear group $i$ is
\begin{equation}
\label{eq:all-particle}
\Phi_i(E)\,\equiv\, E\frac{{\rm d}N_i}{{\rm d}E} \,= \,
\sum_{j=1}^3 a_{i,j}E^{-\gamma_{i,j}}\times exp\left\{-\frac{E}{Z_iR_{c,j}}\right\},
\end{equation}
where $\gamma_{i,j}$ is the integral spectral index for 
nuclear group $i$ in Population $j$.  The spectrum is written here
as $\Phi ={\rm d}N/{\rm d}\ln E$ to simplify the Jacobian for the transition to the spectrum
of nucleons, but also because cosmic-ray spectral measurements are invariably
presented as number of events per logarithmic interval of energy.  Thus the
spectrum of nucleons is
\begin{equation}
\label{eq:nucleons}
\Phi_N(E_N)\,\equiv\, E_N\frac{{\rm d}N}{{\rm d}E_N}\, =E_N\phi(E_N)\,
= \,\sum_{i=1}^5 A_i\times\Phi_i(A_iE_N).
\end{equation}

The all-particle spectrum and the corresponding spectrum of nucleons are
shown side by side in Fig.~\ref{fig:allnucleon}.
Because of the steep spectrum, the contributions of nuclei to
the spectrum of nucleons are suppressed.  Note in particular that
helium, which has a harder spectrum than protons and crosses above
protons in the all-particle spectrum, is always subdominant
in the spectrum of nucleons.  Contributions of heavier nuclei are
even smaller.

\begin{figure}[t]
\centerline{\includegraphics[width=9.cm]{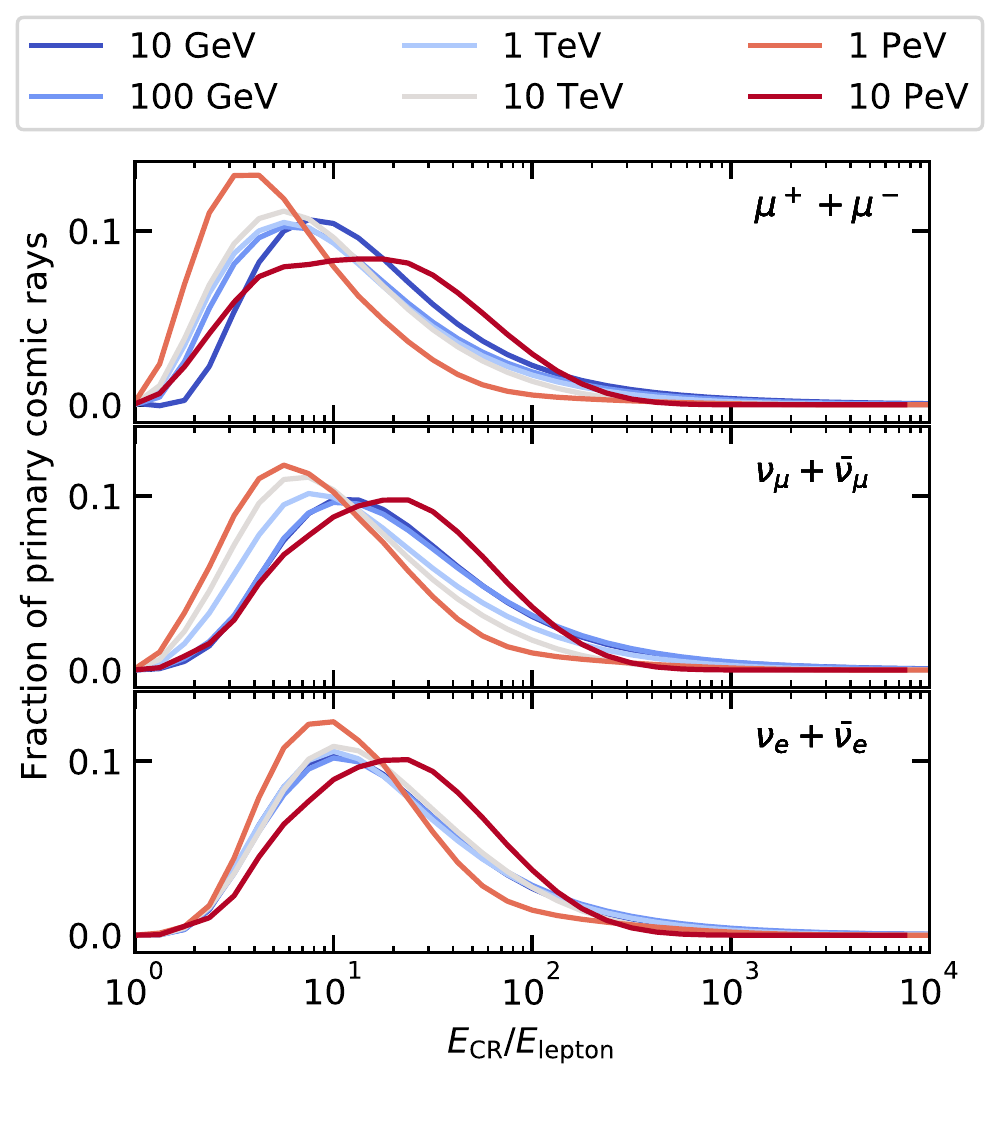}}
\caption{Distributions of primary energy per nucleon for atmospheric leptons.  
(From Ref.\cite{Fedynitch:2018cbl})}
\label{fig:response}
\end{figure}

Figure~\ref{fig:response} from Ref.\cite{Fedynitch:2018cbl} shows the distribution
of primary energies per nucleon ($E_N$) that produce leptons of energy $E_\ell$
as a function of the ratio of the energies.  The distributions reflect the underlying
ratios of $E_N/E_m$ and $E_m/E_\ell$ and the steepness
of the cosmic-ray spectrum (m$=\pi^\pm,\,K^\pm,\,D$).  For $E_\ell\approx 1$~PeV, the steeper spectrum
above the knee shifts the primary energy down.  For $10$~TeV, the prompt
component dominates and the distributions reflect the kinematics
of charm decay for neutrinos and also the unflavored component for muons.
At lower energies, the curves for $\mu$ and $\nu_e$ approximately
scale, while those for $\nu_\mu$ reflect the transition to
dominance of the kaon channel.

\subsection{Direct measurements and Population 1}
Early measurements~\cite{Ahn:2009tb,Ahn:2010gv} with the CREAM balloon calorimeter
were used to determine the parameters $a_{i,1}$ and $\gamma_{i,1}$ of the first
population in H3a.  There are now direct
measurements with spectrometers from PAMELA~\cite{Adriani:2011cu} and 
AMS02~\cite{Aguilar:2015ooa,Aguilar:2015ctt} as well
as a recent measurement by CREAM~\cite{Yoon:2017qjx}.  In their measurements
of protons and helium, PAMELA confirmed what CREAM~\cite{Ahn:2010gv} called a 
``discrepant hardening" of the elemental spectra.  The hardening of the spectrum
above 200 GV rigidity has now been measured by AMS02~\cite{Aguilar:2017hno} for helium,
carbon and oxygen, as well as for protons~\cite{Aguilar:2015ooa}.

\begin{figure}[t]
\centerline{\includegraphics[width=6.3cm]{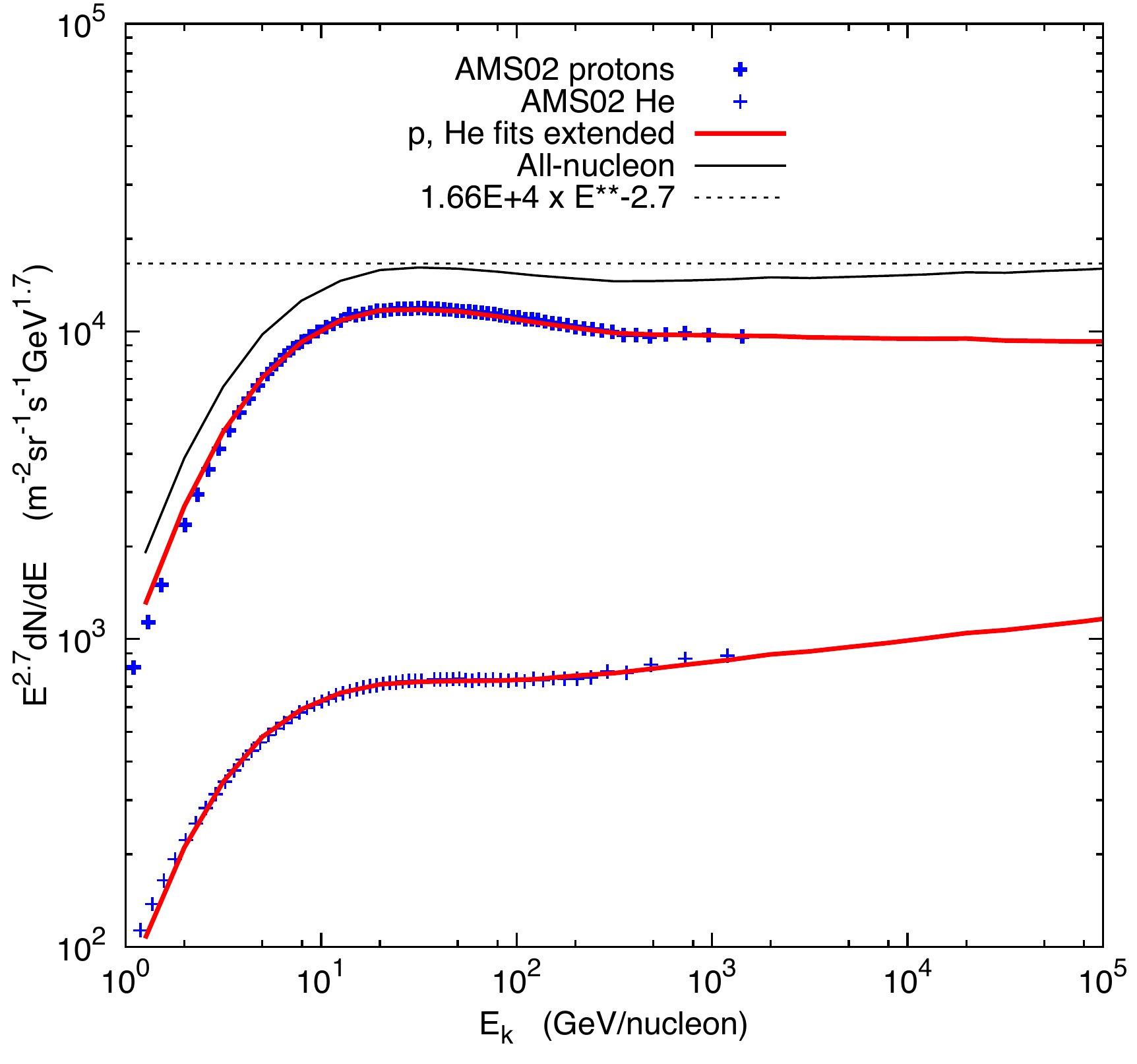}\hspace*{4pt}
\includegraphics[width=6.3cm]{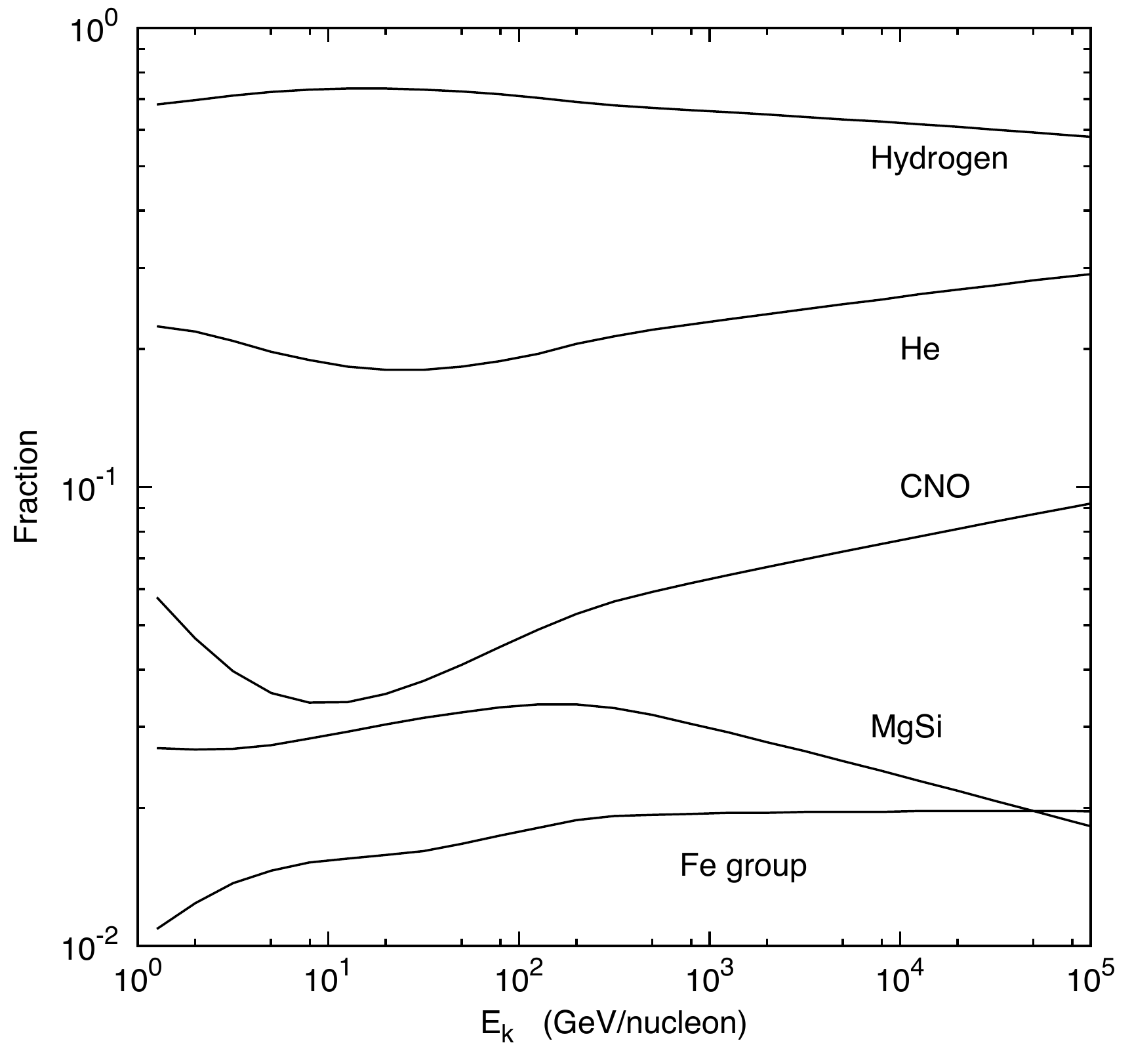}}
  \caption{Left: Extrapolation of AMS02-data for protons and helium to high energy, with
  Helium plotted as number of nuclei per energy per nucleon; Right:
Fractional contributions to the spectrum of nucleons.  Protons and He
  are from AMS02~\cite{Aguilar:2015ooa,Aguilar:2015ctt}; 
  heavier nuclear groups are from Ref.~\cite{Gaisser:2002jj}.}  
     {\label{fig:proton-He}}

\end{figure}

A theoretical interpretation~\cite{Blasi:2012yr,Evoli:2018nmb} is that below $\approx 200$~GV
propagation of cosmic rays in the Galaxy in dominated by turbulence self-generated
by the cosmic rays, which gives a relatively soft spectrum. 
Then there is a transition at high rigidity to a regime in which the propagation is
dominated by external turbulence, for example, turbulence driven by supernova explosions.
 The observed spectral index
is the combination of the injection spectrum at the sources and the energy 
dependence of diffusion, $\gamma = \gamma_s + \delta(R)$.  In this picture,
$\delta(R< 200) > \delta(R> 200)$, and
the high energy behavior can be expected to apply up to the knee.
Fits to AMS02 data for protons~\cite{Aguilar:2015ooa} 
and helium~\cite{Aguilar:2015ctt} (Fig.~\ref{fig:proton-He})
have been used for the first population in a revised H3a model for the extrapolation
to the knee.  
Cosmic-ray spectrum figures in this paper are made 
with this revision.\footnote{
Population 2 is unchanged from Ref.~\cite{Gaisser:2011cc}, and population 3 
is reduced by $30$\%.}

The proton and helium data below $10$~GeV are strongly affected by solar modulation, 
so the fits in Fig.~\ref{fig:proton-He} require parameterizations that
account for the characteristic suppression at low energy.  For high energy,
they also require a form that changes the slope.  A seven parameter form
suggested by Evans \textit{et al.}~\cite{Evans:2016obt} is used:
\begin{equation}
\label{eq:seven-param}
\frac{{\rm d}n}{{\rm d}E_k} = a\left[E_k+b\exp\left(-c\sqrt{E_k}\right)\right]^{-\alpha_1}
\times \left[1+\left(\frac{E_k}{E_c}\right)^s\right]^{\frac{\alpha_1-\alpha_2}{s}}.
\end{equation}
The first factor is the form~\cite{Gaisser:2001jw,Gaisser:2002jj} based on direct 
measurements and used for several of the atmospheric neutrino calculations that 
include solar modulation at low energy\footnote{A useful new treatment 
of solar modulation with parameterizations for the last few solar cycles is the paper of 
Ref.~\cite{Cholis:2015gna}.} and extend to $E_\nu=10$~TeV.  
The second factor is the form originally suggested in~\cite{TerAntonyan:2000hh} 
to make a smooth parameterization of the steepening at the
knee of the spectrum (see also~\cite{Lipari:2017jou}).  Here it provides the
transition from a steeper to a harder power ($\alpha_2<\alpha_1$) 
at a rigidity $\approx 240$~GV.  In Ref.~\cite{Evans:2016obt} Eq.~\ref{eq:seven-param}
was used to fit several proton data sets.  Here it is based specifically on the
AMS02 measurements of protons and helium with the parameters in Table~\ref{tab:7params}.

\begin{table}[ht]\caption{Parameters for Eq.~\ref{eq:seven-param}.}\label{tab:7params}
\centering
\begin{tabular}{cccccccc} \hline
&$a$~(m$^{-2}$sr$^{-1}$s$^{-1}$) & $b$ & $c$ (GeV$^{-\frac{1}{2}}$) & $E_c$ (GeV) & $s$ & $\alpha_1$ & $\alpha_2$ \\ \hline
p & $2.17\times10^4$ & 2.4 & 0.12 & 240. & 3.56 &2.845 & 2.71 \\
He & 800. & 2.0 & 0.33 & 122. & 3.5 & 2.72 & 2.63 \\ \hline
\end{tabular}
\end{table}

\subsection{Neutrino flux from a non-power-law spectrum}
In order to calculate the spectrum of neutrinos over a wide energy
range, it is necessary to take account of deviation of the primary
spectrum from the power-law behaviour assumed for Eq.~\ref{numuflux}.
To do so in the framework of the analytic approach used here, it is 
necessary to use energy-dependent spectrum weighted moments
as described in Ref.~\cite{Gondolo:1995fq}.  For K$^+$ production, 
for example, the generalization of 
$$Z_{pK^+}=\int_0^1\,x^\gamma \frac{{\rm d}n_{K^+}(x)}{{\rm d}x}$$ is
\begin{equation}
\label{eq:EZfactors}
Z_{NK^+}(E)\,=\,\int_E^\infty\,{\rm d}E'\frac{\phi_N(E')}{\phi_N(E)}
\frac{\lambda_N(E)}{\lambda_N(E')}\frac{{\rm d}n_{K^+}(E',E)}{{\rm d}E}.
\end{equation}
As this equation makes clear, this \textit{anzatz} builds in energy dependence
of interaction and production cross sections as well as the deviation from
a power-law spectrum.  However, its implementation requires full knowledge of the 
hadronic production cross sections as a function of beam energy $E^\prime$ for the full
phase space of secondary energy $E$. 

\begin{figure}[t]
\centerline{\includegraphics[width=9.cm]{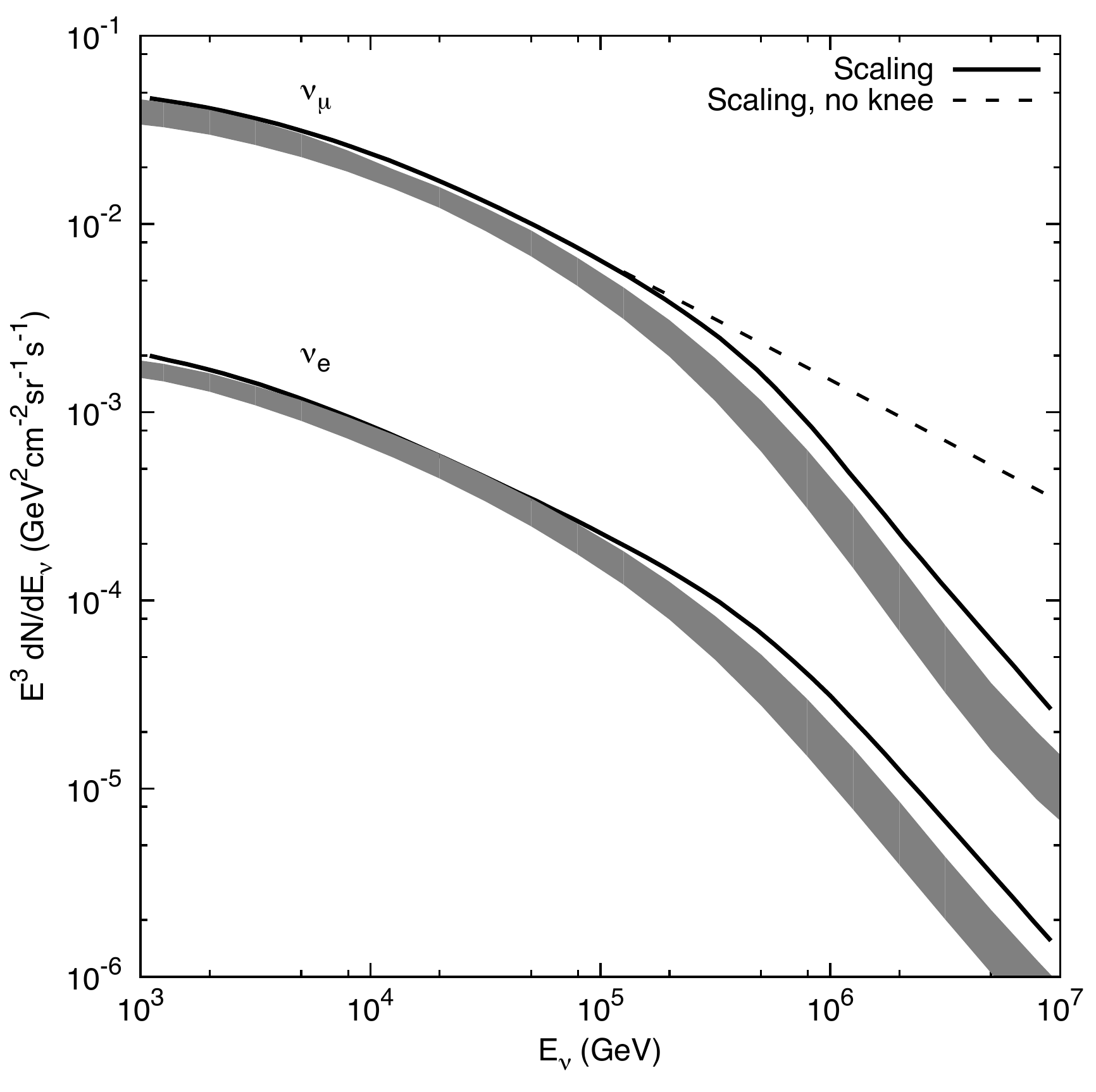}}
\caption{Shaded bands show the range of uncertainties in the angle-averaged neutrino
fluxes for several models of hadronic interactions.  The solid line shows for
comparison the flux from a calculation in which the spectrum weighted moments
are constant and the only source of energy dependence is from the spectrum.  The dashed
line shows the flux for a single power-law spectrum with no knee.}
\label{fig:uncertainties}
\end{figure}

An approximate scheme for handling energy-dependent production cross sections is worked out
in Ref.~\cite{Gaisser:2014eaa}.  It uses two-parameter fits for the production
cross sections~\cite{Gaisser:2001sd} of the form $dn_{ji}/dx=c_{ji}(E)(1-x)^{p_{ji}(E)}\,/\,x$
at closely spaced intervals of energy to obtain the
energy dependence of the $c$ and $p$ parameters.  This procedure is used 
in a subsequent paper~\cite{Gaisser:2016obt}
to compare atmospheric neutrino fluxes for a range of
interaction models and primary spectra.  The main conclusion is that the
variations due to different representations of the knee in the primary
spectrum for a given interaction model are not as large as variations
among different hadronic interaction models for a given primary spectrum.
The uncertainty range for conventional neutrinos from 
several interaction models is shown in Fig.~\ref {fig:uncertainties}.
The shaded bands include Sib2.1~\cite{Ahn:2009wx}, EPOS-LHC~\cite{Pierog:2013ria}, 
QGSJet II-04~\cite{Ostapchenko:2013pia} and Sib 2.3~\cite{Riehn:2015oba}.
The variation is $\approx\pm15$\% from 1 to 30~TeV, increasing to $\approx\pm40$\%
above in the PeV range.  The corresponding ranges for differences in primary
spectra are $\pm10$\% to $\pm15$\%.

These ranges of uncertainties should be compared with an extensive
evaluation of uncertainties in lepton fluxes in Ref.~\cite{Fedynitch:2012fs}.  
An earlier assessment of uncertainties
in fluxes of high-energy atmospheric muons only is Ref.~\cite{Kochanov:2008pt}.

\section{Atmospheric neutrinos with $E_\nu> 10$ TeV}
\label{section:high-energy}
IceCube has discovered a diffuse flux of astrophysical neutrinos from the whole 
sky~\cite{Aartsen:2013jdh,Aartsen:2014gkd} by selecting events that start well inside its 
deep array of optical modules.
This high-energy starting event (HESE) sample includes both tracks initiated by
charged current (CC) interactions of $\nu_\mu$ and
 cascades from $\nu_e$, $\nu_\tau$ and neutral current interactions of all flavors.
 The cascade sample also includes  
some $\nu_\mu$ charged current  events in which most of the energy is given to the
nuclear fragmentation.  The energy threshold is $30$~TeV.
The starting event analysis is extended to lower energy
($\sim$TeV) by progressively increasing the outer veto region~\cite{Aartsen:2014muf}.
Neutrinos of astrophysical origin are also detected~\cite{Aartsen:2016xlq}
above the steeply falling background of atmospheric neutrinos
from analysis of upward tracks, most of which start from CC interactions
of $\nu_\mu$ in the rock and ice outside the detector.
In both cases, it is important to understand the background of
atmospheric neutrinos.

\begin{figure}[t]
\centerline{\includegraphics[width=7.5cm]{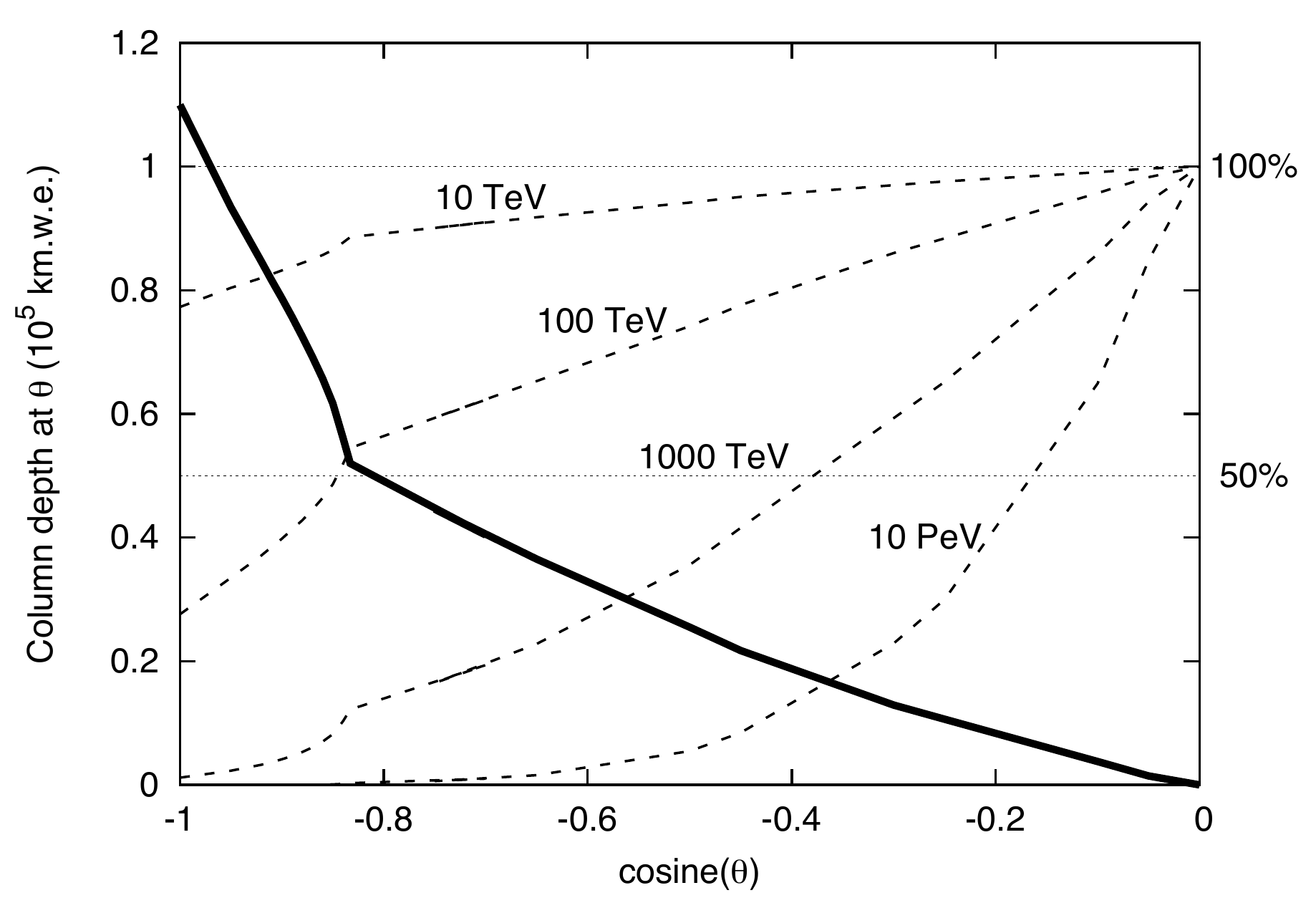}\hspace*{4pt}
\includegraphics[width=6.5cm]{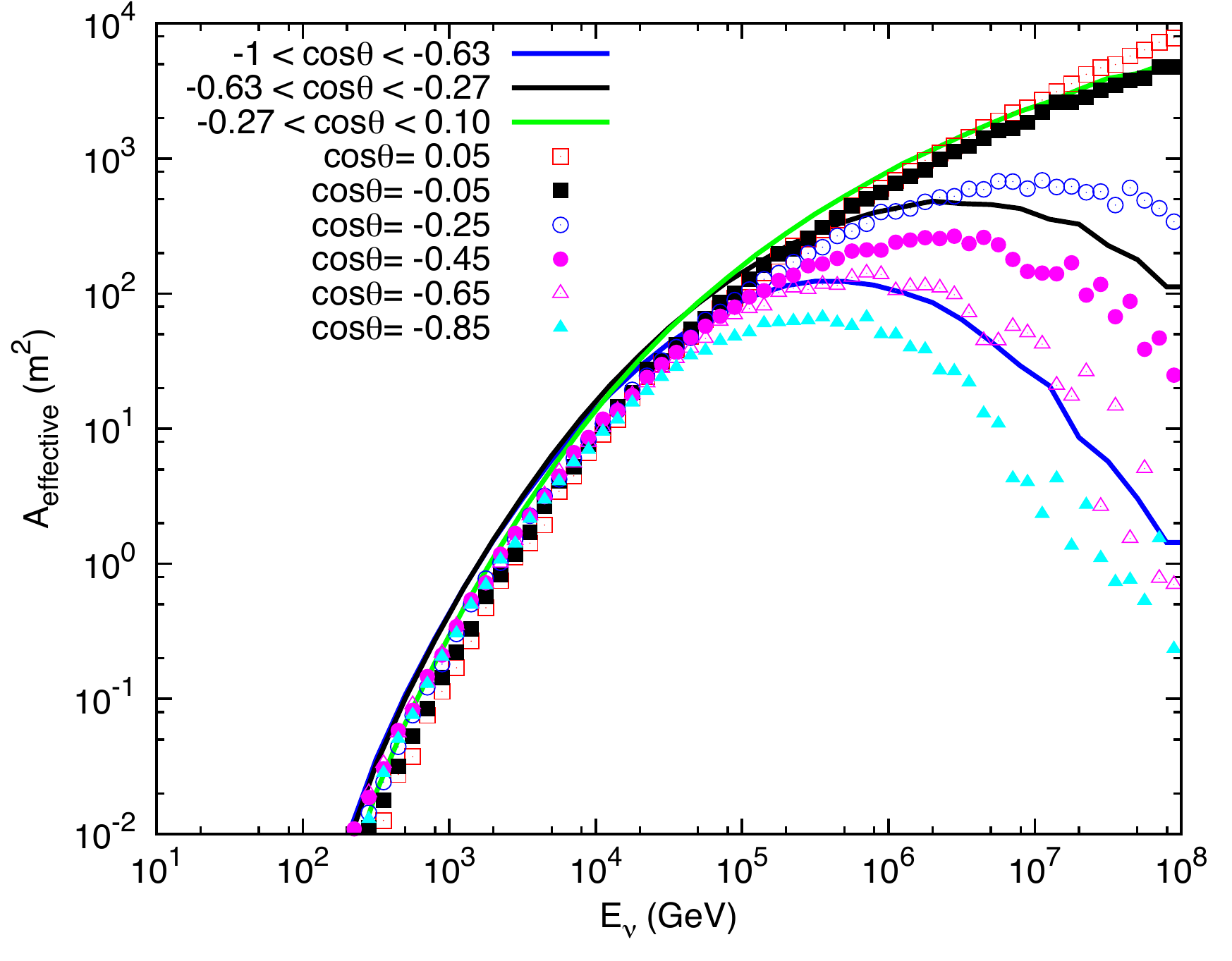}}
  \caption{Left: Diagram showing absorption of muon neutrinos by the Earth; Right:
 Effective area for upward neutrinos in IceCube.  (See text for explanation.)}
     \label{fig:absorption}
\end{figure}

\subsection{Upward neutrino-induced muons}
Selecting upward moving tracks uses the Earth as a filter
to remove all backgrounds except atmospheric $\nu_\mu$.
\subsubsection{Transparency of the Earth}
In this case it is necessary to understand the dependence
of absorption by the Earth as a function of zenith angle
and neutrino energy.  Figure~\ref{fig:absorption}~(left) shows
how the Earth shadows upward-moving neutrinos.  The solid line shows 
the thickness of the Earth as a function of direction~\cite{Gandhi:1995tf}.
The dashed lines show the fraction of neutrinos absorbed for
selected neutrino energies.  Absorption affects both astrophysical
and atmospheric neutrinos.  It starts to become
important for $E_\nu\approx 10$~TeV.  For higher energies,
increasing absorption limits trajectories to near the
horizon.  For conventional atmospheric neutrinos, absorption enhances the
intrinsic preference for large angles from the $\sec\theta$ effect.

\subsubsection{Effective areas for upward $\nu_\mu$}
 The effective area is
a detector-dependent quantity that connects a flux of neutrinos
with the observed signal.  For upward muon neutrinos 
$\phi_\nu \times A_{\rm eff}$ gives the observed
rate of neutrino-induced muons. 
The effective area for this case is given by
\begin{equation}\label{eq:nuAeff}
A_{\rm eff}(E_\nu,\theta)\,=\,\epsilon(E_{\rm th},\theta)\,
A(\theta)\,P_\nu(E_\nu,E_\mu>E_{\rm th})
\,\exp\{-(1-\zeta)\sigma_\nu(E_\nu)N_AX(\theta)\}.
\end{equation}
Effects of absorption in the Earth are reflected in the
exponential absorption factor where
$X(\theta)$ is the slant depth through the Earth in g/cm$^2$,
$N_A$ is the number of nucleons per gram, $\sigma_\nu$ the 
total neutrino cross section and $\zeta$ accounts for 
energy loss and regeneration in NC interactions.~\cite{Ingelman:1996mj}.
The efficiency of the detector $\epsilon(E_{\rm th},\theta)$ for measuring a muon above
some threshold energy from direction $\theta$ multiplied by
its projected area $A(\theta)$ sets the scale.  The central
factor in Eq.~\ref{eq:nuAeff}, $P(E_\nu,E_\mu>E_{\rm th})$,
is the probability that a muon neutrino
(or antineutrino) of energy $E_\nu$ headed toward the detector will
produce a muon that reaches it with $E_\mu>E_{\rm th}$.\footnote{When
$\tau$ neutrinos are involved, the survival probability has also to
include regeneration via decay of $\tau$ leptons~\cite{GonzalezGarcia:2005xw},
for example for astrophysical $\nu_\tau$~\cite{Halzen:1998be} and for
atmospheric neutrinos with oscillations, including non-standard oscillation
scenarios~\cite{Collin:2016aqd}.  Regeneartion effects are included in Ref.\cite{Delgado:2016abc}.}

Effective areas for the measurement
of diffuse, upward muon neutrinos by IceCube~\cite{Aartsen:2016xlq} are shown by
the full lines in the right panel of Fig.~\ref{fig:absorption} for three ranges of zenith angle.
The broken lines show the effective area for upward neutrinos in
the IceCube point source search.\cite{IceCube:2016abc}

\subsection{Starting events and neutrino self-veto}
\begin{figure}[t]
\centerline{\includegraphics[width=6.5cm]{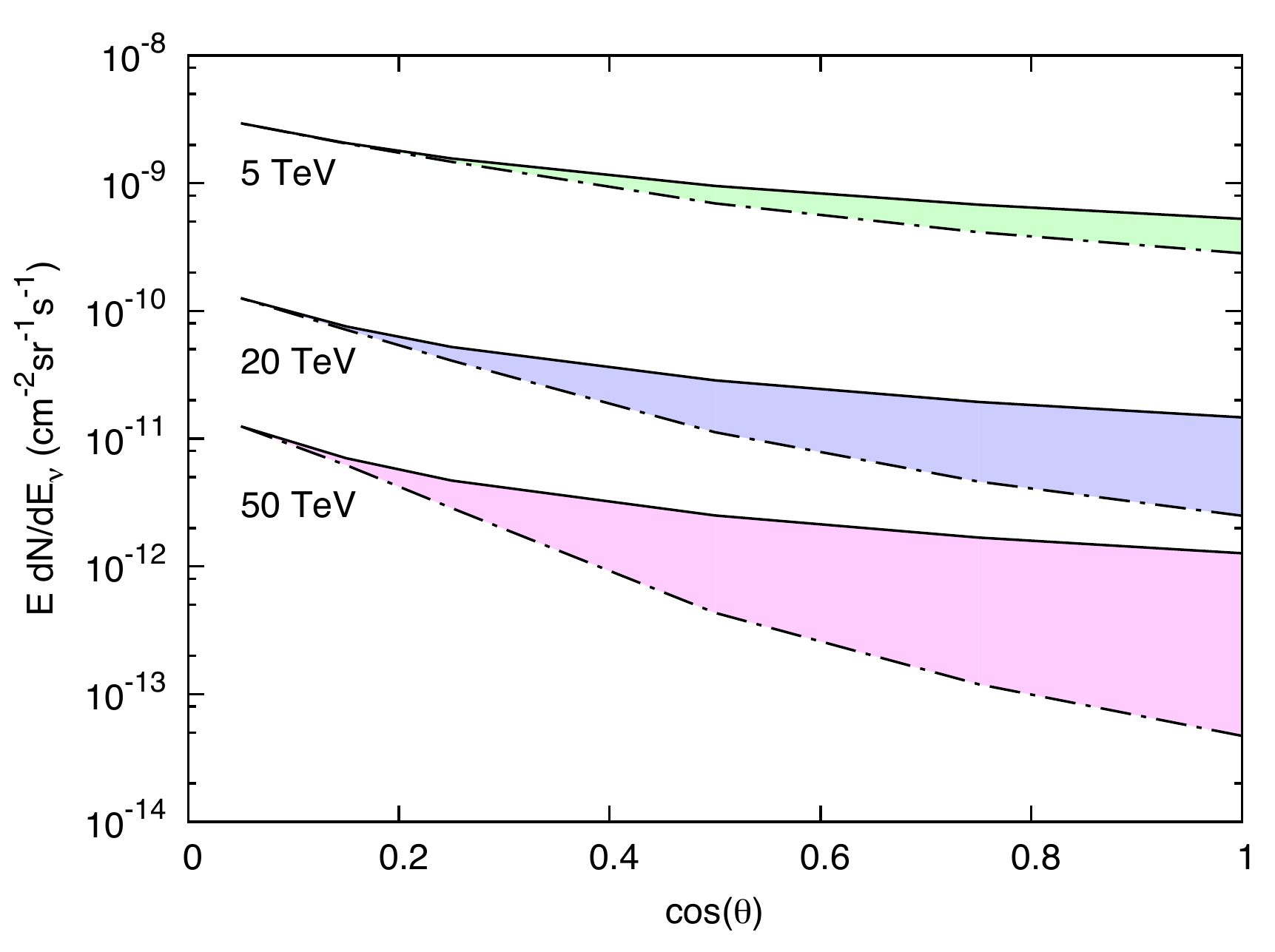}\hspace*{4pt}
\includegraphics[width=6.5cm]{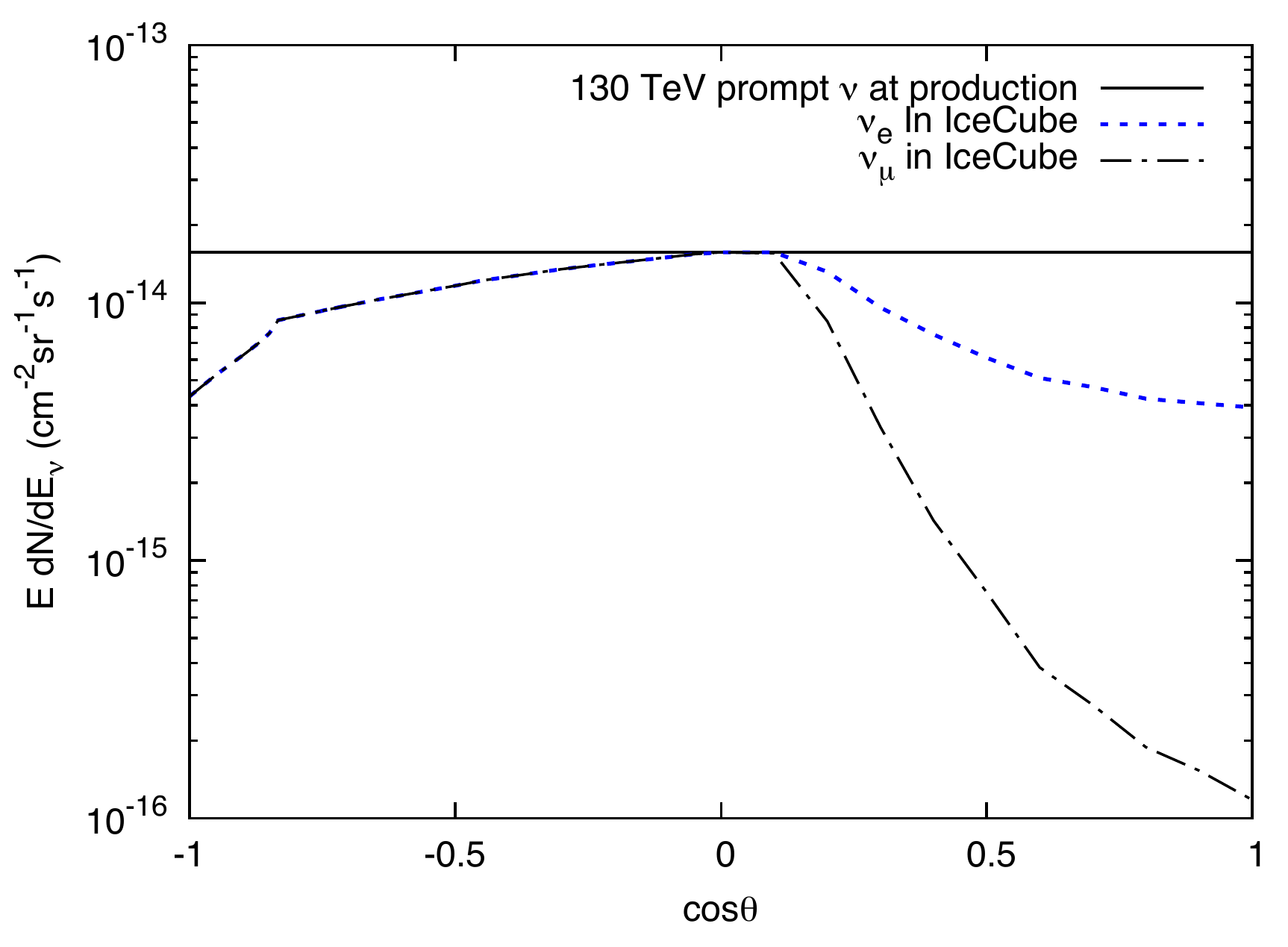}}
  \caption{Left: Angular dependence of fluxes of conventional muon neutrinos at three energies,
  with shaded regions showing the effect of the neutrino self-veto; Right:
Angular dependence of prompt neutrinos at 130 TeV, with full lines
  showing the distributions at production and broken lines 
  the flux at the center of IceCube after absorption by the Earth ($\cos\theta< 0.$) and
  after applying the passing fraction of Ref.\cite{Arguelles:2018awr}.}  
     \label{fig:fractional}
\end{figure}

The stating event analyses in IceCube~~\cite{Aartsen:2013jdh,Aartsen:2014gkd} depend 
on a veto region to exclude entering muons and select only neutrino interaction 
vertices well inside the detector.  This has an important effect on the
atmospheric neutrino background from above the detector: neutrinos
accompanied by muons at the detector will be classified as muons and
therefore excluded from the neutrino sample.  
Only those atmospheric neutrinos that are
not accompanied by a muon at the detector will be included.  
This self-veto effect increases
with energy because the likelihood that a muon produced in the same event has
sufficient energy to reach the deep detector increases.
The flux of atmospheric neutrinos
at the detector is then expressed as
\begin{equation}\label{eq:passing}
\phi_{\nu,{\rm det}}(E_\nu,\theta) = {\cal P}_i(E_\nu,E_\mu^{\rm min},\theta)\phi_\nu(E_\nu,\theta),
\end{equation}
where the passing rate ${\cal P}_i$ depends on neutrino flavor ($i$), energy and angle
and on the depth of the detector through $E_\mu^{\rm min}$.
The passing rate is lowest at the zenith and increases to unity at the horizon
as the energy required of the muon to reach the detector increases because of the
larger slant depth.  This introduces an additional angular dependence to the
atmospheric neutrino distributions at the detector, favoring large angle more
and more as energy increases, as shown in Fig.~\ref{fig:fractional}~(left).  
Unlike the increased enhancement toward the
horizon in the case of upward $\nu_\mu$, which affects both atmospheric and
astrophysical neutrinos, here it is only the atmospheric neutrinos that are affected.
Thus the energy and angular dependence of ${\cal P}_i$ in the case of neutrinos
from above provide additional discrimination power between astrophysical and atmospheric
neutrinos.

For a given energy and angle the passing fraction is much smaller for $\nu_\mu$
than for $\nu_e$ because of the muon that is produced in the same decay as
the $\nu_\mu$.  The contribution to the passing fraction from the sibling
muon was calculated in Ref.~\cite{Schonert:2008is} in the analytic framework
of Ref.~\cite{Gaisser:2016uoy} by adding the requirement $E_\mu>E_\mu^{\rm min}$
to the integrals of pion and kaon decay in the neutrino production spectrum.
The calculation illustrated the point that the passing fraction increases
with the depth of the detector because $E_\mu^{\rm min}$ is larger.

To generalize the calculation of the passing rate to include the case in which
the vetoing muon is produced elsewhere in the shower requires integrating over
the properly weighted distribution of primary cosmic-rays capable of producing
the target neutrino.  For each step of the integral, the probability that
a cosmic-ray of the given energy will produce a muon with $E_\mu>E_\mu^{\rm min}$
is accumulated.  In this way a generalized passing fraction is calculated
that applies also to electron neutrinos and which, when combined with the analytic estimate,
slightly reduces the passing fraction for muon neutrinos.  The generalized self-veto
was first implemented in Ref.~\cite{Gaisser:2014bja} using the Elbert 
formula~\cite{Elbert:1979gz,Elbert:1979abc} to evaluate the muons produced elsewhere
in the shower.  The original parameters of the Elbert formula were tuned to improve
agreement with a limited set of Monte Carlo calculations.  This passing fraction was
applied to the IceCube $>1$~TeV starting event analysis~\cite{Aartsen:2014muf}
and to the four-~\cite{Aartsen:2015zva} and six-year~\cite{Aartsen:2017mau} HESE analyses 
with $E_\mu^{\rm min}$ calculated so that the vetoing muon has $1$~TeV at the detector.

A recent paper~\cite{Arguelles:2018awr} unifies and improves the accuracy of the
passing fraction in several ways.  The MCEq~\cite{Fedynitch:2017abc} program is used to evaluate muons
produced elsewhere in the shower, taking account of the energy removed from
the primary cosmic-ray by the neutrino.  Fluctuations in muon energy loss are accounted
for, which smooths out the shoulder in the passing fraction for the sibling muon
visible in the plots of Ref.~\cite{Schonert:2008is}.  The numerical calculations
are compared to an application specific Monte Carlo~\cite{Jero:2016abf} designed
to evaluate neutrino fluxes efficiently.  The comparison confirms an interesting feature
of the calculation, which is that passing fractions for conventional anti-neutrinos
are slightly lower than for conventional neutrinos.  This is attributed to the fact
that negative mesons, and therefore their decay products including anti-neutrinos, 
on average carry a smaller fraction of the
primary energy, leaving more energy in the rest of the shower to provide a vetoing muon. 
The passing rates are tabulated over a 
range of energies and directions extending to very low values for high energy
and directions near the vertical.  For the purpose of evaluating the shape
and level of the surviving atmospheric neutrinos, passing rates much less than
the fractional uncertainties from treatment of hadronic interactions and the primary
spectrum are equivalent to zero.  In the case of a very high energy astrophysical
candidate, however, it may be important to have a precise knowledge of a very small
passing probability.

The combination of absorption in the Earth for upward events
and the self-veto effect for downward events provides a distinctive energy-dependent signature
for prompt neutrinos, as illustrated in the right panel of Fig.~\ref{fig:fractional}.  The energy chosen in
the plot is the crossover region between prompt and atmospheric electron neutrinos.
The challenge is that the expected rates of prompt $\nu_e$ 
in the HESE analysis are low~\cite{Gaisser:2014eaa}.

\section{Concluding comment}
The analytic/numerical approach illustrated in this chapter remains useful
for insight into the physics and phenomenology of atmospheric lepton fluxes.
It can also provide the basis for quantitative estimates of their uncertainties.


\section*{Acknowledgments}

\noindent
I am grateful to Anatoli Fedynitch and Maria Vittoria Garzelli for helpful comments
on this paper and to
Marjon Moulai, K. Okumura and Tianlu Yuan for relevant discussions.  
This work is supported in part by the U.S. National Science
Foundation (PHYS:1505990), by the U.S. Department of Energy (DE-SC0013880), by the Bartol Research 
Institute of the University of Delaware and by the Munich Institute 
for Astro- and Particle Physics (MIAPP) of the DFG cluster of excellence ``Origin and Structure of the Universe".

\chapter*{References}
{\footnotesize
\bibliography{Gaisser}

%

}

\end{document}